\documentclass[fleqn,usenatbib]{mnras}

\usepackage{newtxtext,newtxmath}

\usepackage[T1]{fontenc}

\DeclareRobustCommand{\VAN}[3]{#2}
\let\VANthebibliography\thebibliography
\def\thebibliography{\DeclareRobustCommand{\VAN}[3]{##3}\VANthebibliography}

\usepackage{graphicx}	
\usepackage{amsmath}	

\title[Multi-Epoch Machine Learning 2]{Multi-Epoch Machine Learning 2: Identifying physical drivers of galaxy properties in simulations}

\author[Robert McGibbon \& Sadegh Khochfar]{
Robert J. McGibbon,$^{1}$\thanks{E-mail: rob.mcgibbon@ed.ac.uk}
Sadegh Khochfar$^{1}$
\\
$^{1}$Institute for Astronomy, University of Edinburgh, Royal Observatory, Edinburgh EH9 3HJ
}

\date{Accepted XXX. Received YYY; in original form ZZZ}
\pubyear{2023}

\begin{document}
\label{firstpage}
\pagerange{\pageref{firstpage}--\pageref{lastpage}}
\maketitle

\begin{abstract}
Using a novel machine learning method, we investigate the buildup of galaxy properties in different simulations, and in various environments within a single simulation. The aim of this work is to show the power of this approach at identifying the physical drivers of galaxy properties within simulations. We compare how the stellar mass is dependent on the value of other galaxy and halo properties at different points in time by examining the feature importance values of a machine learning model. By training the model on IllustrisTNG we show that stars are produced at earlier times in higher density regions of the universe than they are in low density regions. We also apply the technique to the Illustris, EAGLE, and CAMELS simulations. We find that stellar mass is built up in a similar way in EAGLE and IllustrisTNG, but significantly differently in the original Illustris, suggesting that subgrid model physics is more important than the choice of hydrodynamics method. These differences are driven by the efficiency of supernova feedback. Applying principal component analysis to the CAMELS simulations allows us to identify a component associated with the importance of a halo's gravitational potential and another component representing the time at which galaxies form. We discover that the speed of galactic winds is a more critical subgrid parameter than the total energy per unit star formation. Finally we find that the Simba black hole feedback model has a larger effect on galaxy formation than the IllustrisTNG black hole feedback model.
\end{abstract}




\section{Introduction}
\label{sec:introduction}

Hydrodynamical cosmological simulations have become key tools for helping us to understand both cosmology and galaxy formation. 
These simulations include baryons alongside dark matter, and incorporate relevant physical processes such as gas cooling, star formation, and feedback.
Prominent examples of cosmological simulations include Illustris \citep{illustris_1, illustris_2, illustris_3, illustris_4}, IllustrisTNG \citep{tng_1, tng_2, tng_3, tng_4, tng_5}, Simba \citep{simba}, EAGLE \citep{eagle_1, eagle_2}, HorizonAGN \citep{horizon_agn_1}, and FiBY \citep{fiby_1}. 
They have been successful at reproducing a large number of observables, such as the column density distribution of the Lyman-$\alpha$ forest \citep{lyman-alpha_forest}, stellar mass and luminosity functions \citep{eagle_1, tng_2, fiby_2}, galaxy clustering \citep{tng_1}, the galaxy color bimodality \citep{horizon_agn_2, eagle_colours, tng_4}, and properties of the circumgalactic medium \citep{appleby_cgm}.
For recent reviews of cosmological simulations see \cite{dave_review} and \cite{vogelsberger_review}.

There are various methods for simulating gas, including particle based methods \citep[e.g.][]{gadget_2}, and grid based approaches, using both structured and unstructured meshes \citep[e.g.][]{arepo, 2023MNRAS.518.4401M}, with some codes utilising adaptive mesh refinement \citep[e.g.][]{enzo}.
Each comes with its own strengths and weakness, and numerical effects from the implementations of the various methods can affect the hydrodynamics of the gas in different ways.
More variation in simulations comes from the fact that many of the relevant physical processes that need to be modelled occur below the typical resolution limits of cosmological simulations.
Therefore various 'sub-grid physics' prescriptions are employed to model them. 
There are many different implementations of different subgrid models, and each model tends to have a number of tunable parameters.

\begin{figure}
	\includegraphics[width=\columnwidth]{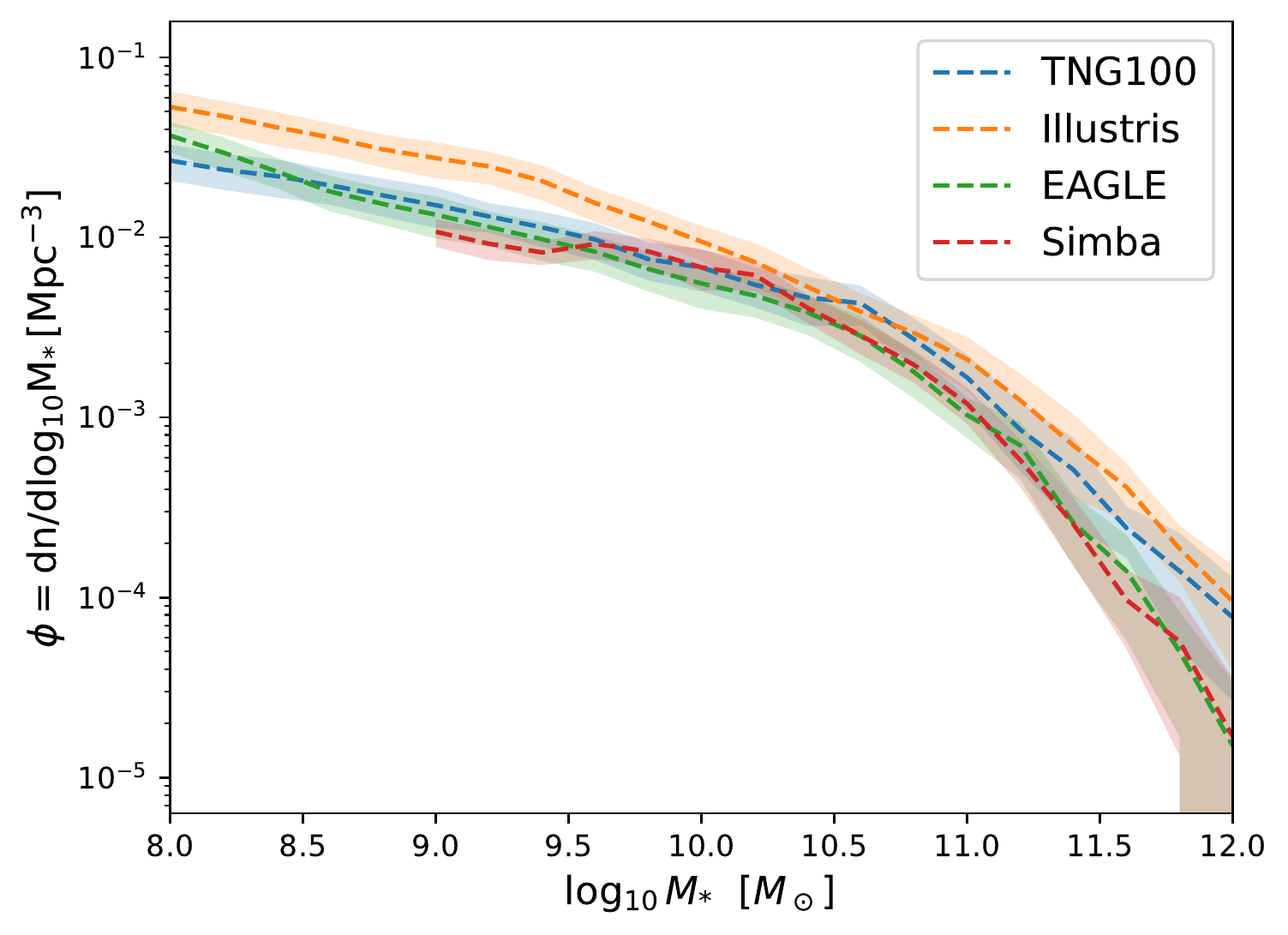}
    \caption{Stellar mass functions from the Illustris, TNG100, EAGLE, and Simba simulations. The shaded area corresponds to the spread computed from jackknife re-sampling over eight simulation sub-octants.}
    \label{fig:stellar_mass_function}
\end{figure}

Even with these variations in modelling, subgrid models from different simulations can be tuned to reproduce observables. 
An example of this is given in Figure \ref{fig:stellar_mass_function}, which shows the $z=0$ stellar mass function from a number of simulations.
The results from Simba and EAGLE agree over the full mass range of galaxies they resolve. IllustrisTNG is also consistent with EAGLE, apart from at the highest masses. The original Illustris run does not show as good agreement, but displays the same general behaviour as the other simulations.
Despite the similar stellar populations at $z=0$, the way in which galaxies build up their stellar mass can differ greatly between simulations, which is what the technique presented in this paper is able to distinguish.

In this work we focus on implementations of stellar and black hole feedback.
In order to regulate star formation, stellar feedback must effectively generate galactic-scale outflows to remove gas from galaxies. 
Various sub-resolution techniques can be used to achieve this, each with different methods for transferring energy and momentum, primarily from supernova explosions, to the surrounding gas. This energy can be added thermally 
or kinetically. 
Black hole feedback is related to a number of observational phenomena such as relativistic jets and the black hole mass and bulge mass relation.
This feedback is typically divided into two modes: quasar mode, which is associated with the radiatively efficient mode of black hole growth, and jet mode, which is associated with highly-collimated jets that possess enough energy to counteract cooling losses. These modes are implemented differently in simulations, but are generally accomplished through momentum or energy injection.

As cosmological simulations output large volumes of data, machine learning can be a useful technique for gaining understanding of the processes occurring within the simulation. These kinds of methods can be used for obtaining insights into both dark-matter-only and hydrodynamical simulations. We highlight a number of recent examples below.

Using N-body simulations \cite{lucie-smith_1} trained a decision-tree based model to predict the final mass of halos by using the simulation initial conditions as input, and examined their model to determine the importance of the tidal shear field in establishing halo mass.
In \cite{lucie-smith_2} the authors predicted the density profiles of halos, and included the mass accretion history of the halo as model input. They showed how the feature importance from the model could be interpreted in terms of physical timescales.

By training neural networks on hydrodynamical cosmological simulations, and then applying symbolic regression, \cite{virial} were able to recover a version of the virial theorem.
\cite{accreted_stellar_mass} used random forests to learn which galaxy properties are most useful for determining the fraction of accreted stellar mass, and compared the feature importance values for high mass and low mass galaxy samples.
\cite{ergo-ml} showed that by using invertible neural networks it is possible to gain information about the properties that are most predictive of the time of a galaxies' last major merger.
\cite{assembly_bias} used saliency maps from a convolutional neural network to show how the H\textsc{i} content of halos exhibits a strong dependence on their local environment.
\cite{gnn_halo_mass} trained a graph neural network to predict the mass of a halo based on its host galaxies, and examined which galaxy properties were the most predictive.

These types of machine learning methods are not only restricted for use with simulations, but can also be applied directly to observations to advance our understanding of galaxy formation processes.
\cite{rf_observation_1} and \cite{rf_observation_2} both used the feature importance from trained random forest models to examine the most important parameters for predicting whether a galaxy is quenched.
\cite{bpt_scatter} investigated which physical properties are most predictive for determining the position of galaxies on BPT diagrams by using neural networks and random forests, and \cite{gama_som} used self-organizing maps to help analyse galaxy bimodalities.

One common application of machine learning to cosmological simulations has been to train a model to predict the baryonic properties of dark matter only halos \citep[e.g.][]{galaxynet_painting, shap_painting, sparse_1_painting, de-santi_painting, stiskalek_painting, harry_painting, gnn_painting, croc_painting, de-andres_painting, sparse_2_painting}.
In \cite{multi_epoch_1} the authors introduced a new method for making these predictions, which used the properties of the subhalo over a wide range of times as model input.
We showed how by examining the importance of the various input features, it was possible to extract insights into the physical processes occurring within the simulations used to train the models, specifically information on when was the most important time for determining the values of the different physical properties.
In this work we expand on the method introduced in \cite{multi_epoch_1}.
We now use baryonic features as input to the model, rather than just halo properties, and study the resulting feature importance plots.
The purpose of this is to gain insight into the different formation processes leading to distinct populations within a given simulation and also that occur within different simulations.
Using baryonic features as inputs allows us to more easily examine the impact of feedback than if we only considered halo properties.

The remainder of this paper is organized as follows. 
In Section \ref{sec:methods} we give an overview of the simulations used in this work, focusing on the differences in their feedback subgrid model implementations. We also describe our machine learning method.
In Section \ref{sec:subsamples} we test the robustness of our method and show how it can highlight difference in galaxy populations by applying it to the IllustrisTNG simulation suite.
In Section \ref{sec:compare_sim} we look at the insights we can gain when comparing different simulations.
In Section \ref{sec:camels_results} we apply our method to the CAMELS simulation suite, and show how the feature importance changes when subgrid model parameters are varied.
We discuss how our results relate to existing literature and summarize our findings in Section \ref{sec:conclusions}.

\section{Methods}
\label{sec:methods}

\subsection{Simulations}
\label{sec:simulations}

In this subsection we summarise the hydrodynamical cosmological simulations used in this work.
Each simulation includes all significant physical processes required to track the evolution of dark matter, cosmic gas, luminous stars, and supermassive blackholes (SMBHs) from high redshifts ($z\sim 100$) to the present day $z=0$.
We will focus on the different subgrid implementations of supernova and black hole feedback.
We quote the dark matter particle mass, $m_{\text{DM}}$, as a measure of the resolution of the simulation rather than attempting to directly compare the mass resolution of different baryonic elements.
For all simulations the halos are first located using the friend-of-friends algorithm (\textsc{FoF}; \cite{fof}), then substructure is identified using the \textsc{SubFind} subhalo finder \citep{subfind}.
To avoid poorly resolved objects we only consider subhalos with $10^{8.5} < M_* < 10^{12}$.

\subsubsection{Illustris}
\label{sec:illustris}
Illustris\footnote{\href{https://www.illustris-project.org/}{illustris-project.org/}} \citep{illustris_1, illustris_2, illustris_3, illustris_4} is run with the moving mesh code \textsc{Arepo} \citep{arepo}. 
The simulation adopts a WMAP-9 \citep{wmap9} consistent cosmology, and merger trees are constructed using the \textsc{SubLink} algorithm \citep{sublink}.
The box size is $(75 \,\,h^{-1}{\rm Mpc})^3$, with $m_{\text{DM}} = 6.3 \times 10^6 M_{\sun}$. 

Feedback associated with star formation is assumed to drive galactic scale outflows.
The generated winds have a velocity scaled to the local dark matter velocity dispersion.
The mass loading factor of the wind is calculated using the desired wind speed and available supernova energy.
The direction of the wind is determined by the parent gas cell in such a way that wind particles are ejected preferentially along the rotation axis of spinning objects.
SMBH feedback occurs in two different modes.
If the Eddington ratio is below 0.05, a radio-mode model injects highly bursty thermal energy into large, $\sim$50 kpc ‘bubbles’ which are displaced away from the central galaxy.
Above this accretion rate, a quasar-mode model injects thermal energy into the immediately surrounding gas.

\subsubsection{IllustrisTNG}
\label{sec:tng}

The IllustrisTNG suite\footnote{\href{https://www.tng-project.org/}{tng-project.org/}} \citep{tng_1, tng_2, tng_3, tng_4, tng_5} is an update to Illustris simulation. 
It is also run using the \textsc{Arepo} code \citep{arepo}, but a notable addition is the inclusion of magnetic fields.
Cosmological parameters are set to the Planck 2015 values \citep{planck_2015}.
There are 3 different box sizes, each with its own resolutions.
Comparable with the original Illustris simulation, TNG100 has a box size of $(75 \,\,h^{-1}{\rm Mpc})^3$ and dark matter particles have $m_{\text{DM}} = 7.5 \times 10^6 M_{\sun}$.
TNG100 uses the same initial conditions as Illustris, although they have been adjusted for the updated cosmology.
TNG300 has a box size of $(205 \,\,h^{-1}{\rm Mpc})^3$ and $m_{\text{DM}} = 5.9 \times 10^7 M_{\sun}$, while TNG50 has a box size of $(35 \,\,h^{-1}{\rm Mpc})^3$ and $m_{\text{DM}} = 4.5 \times 10^5 M_{\sun}$.
Two sets of merger trees are available: one generated using \textsc{Sublink} \citep{sublink}, the second created by \textsc{LHaloTree} \citep{lhalotree}.
 
The TNG model for stellar feedback is based on the Illustris model, with some modifications.
Winds are now ejected isotropically, although will still naturally propagate along the direction of least resistance.
The wind velocity is now redshift-dependent, and a wind velocity floor is also introduced.
The result of these changes is that stellar feedback in the TNG is more effective at suppressing star formation.
The TNG also features two modes of SMBH feedback, with the mode being dependent on whether the Eddington ratio is above a critical value. For the TNG this critical value increases with the mass of the black hole. The high accretion thermal mode is the same as Illustris, but for low accretion rates a kinetic mode is used which adds momentum to neighbouring gas cells.
For more details regarding the specific implementation of the IllustisTNG simulations, including all relevant subgrid models, we refer the reader to \cite{tng_implementation_1} and \cite{tng_implementation_2}.

\subsubsection{EAGLE}
EAGLE\footnote{\href{http://icc.dur.ac.uk/Eagle}{icc.dur.ac.uk/Eagle};   \href{https://eagle.strw.leidenuniv.nl}{eagle.strw.leidenuniv.nl}} \citep{eagle_1, eagle_2} is a suite of cosmological simulations run with the smoothed particle hydrodynamics code \textsc{GADGET-3} \citep{gadget_2} using the \textsc{Anarchy} scheme.
It adopts a Planck 2013 cosmology \citep{planck_2013}.
Merger trees are built using the \textsc{D-Trees} algorithm \citep{d-trees}.
The fiducial EAGLE simulation, named Ref-L100N1504, has a box size of $(100 \,\,{\rm Mpc})^3$, and dark matter particles have $m_{\text{DM}} = 9.7 \times 10^6 M_{\sun}$.

Supernova feedback in EAGLE is implemented by injecting thermal energy in a stochastic manner to nearby particles. The energy injected per unit stellar mass varies based on the metallicity and density of the interstellar medium (ISM).
SMBH feedback in EAGLE is achieved using a single mode of feedback that operates at any Eddington ratio, in contrast to the dual modes of Illustris, TNG, and Simba. Feedback energy is stored until it is sufficient to heat the surrounding particles by $\Delta T = 10^{7.5} \rm{K}$ and then is stochastically injected as thermal energy. This `pulsed' nature of the thermal feedback prevents the energy being immediately radiated away and offsets cooling. This makes it more efficient at quenching the galaxy than the corresponding thermal mode in Illustris and TNG.

Alongside the fiducial EAGLE simulation, we consider some variants run with the same resolution, but differing subgrid models.
We utilise the FBconst and FBZ simulations, which are described in \cite{eagle_3}. 
Both of these runs have been calibrated to match the observed stellar mass function.
The FBconst model injects into the ISM a fixed amount of energy per unit stellar mass formed.
For the FBZ simulation the energy associated with supernova feedback depends on the ISM metallicity, but unlike the fiducial model the energy is independent of gas density.
We also examine the NoAGN run, which has the same subgrid models as the reference EAGLE simulation, but does not include black holes.

\begin{table*}
    \centering
    \caption{The parameters that are varied for each run of the CAMELS simulations. \textbf{Min} and \textbf{Max} give the minimum and maximum values that the parameters take on. \textbf{log scale} indicates whether the values are sampled linearly, or if they are varied with a logarithmic scale. \textbf{IllustrisTNG effect} (\textbf{Simba effect}) gives information about how the value of the parameter changes the feedback models within IllustrisTNG (Simba). For more details about the parameters see \protect\cite{camels_2}. }
    \begin{tabular}{|c|c|c|c|c|c|}
        \textbf{Parameter} & \textbf{Min} & \textbf{Max} & \textbf{log scale} & \textbf{IllustrisTNG effect}                 & \textbf{Simba effect}
        \\ \hline \hline
        $\Omega_m$         & 0.1          & 0.5          & No                 & Initial conditions                           & Initial conditions 
        \\ \hline
        $\sigma_8$         & 0.6          & 1            & No                 & Initial conditions                           & Initial conditions 
        \\ \hline
        $A_{SN1}$          & 0.25         & 4            & Yes                & Energy output per unit star formation        & Wind mass outflow rate per unit star formation
        \\ \hline
        $A_{AGN1}$         & 0.25         & 4            & Yes                & Prefactor for power injected in kinetic mode & Prefactor for momentum flux of outflows
        \\ \hline
        $A_{SN2}$          & 0.5          & 2            & Yes                & Speed of galactic winds                      & Speed of galactic winds 
        \\ \hline
        $A_{AGN2}$         & 0.5          & 2            & Yes                & Burstiness and temperature                   & Speed of jets
        \\ \hline
    \end{tabular}
    
    \label{table:camels_params}
\end{table*}

\subsubsection{CAMELS simulations}
\label{sec:camels}
The CAMELS project\footnote{\href{https://camels.readthedocs.io/}{camels.readthedocs.io/}} \citep{camels_1, camels_2} contains two different suites of state-of-the-art hydrodynamic simulations. 

The simulations in the first suite have been run with the \textsc{Arepo} code \citep{arepo} and employ the same subgrid physics model as the IllustrisTNG simulations. See section \ref{sec:tng} for more details on the subgrid models for this suite.

The simulations in the second suite have been run with the \textsc{GIZMO} code \citep{gizmo} and employ the same subgrid physics model as the Simba simulation \citep{simba}, which built on its precursor MUFASA \citep{mufasa} with the addition of supermassive black hole growth and feedback \citep{simba_bh_model}.
Star formation in the Simba model drives winds similar to those found in the TNG. The mass loading factor is scaled by redshift, and is constant for low mass galaxies.
The Simba model has two SMBH kinetic feedback modes.
At high Eddington ratios SMBHs drive multi-phase winds at velocities of $\sim 10^3 \;{\rm km}\,{\rm s}^{-1}$.
At low Eddington ratios gas is heated to the virial temperature of the halo and ejected at velocities of $\sim 10^4 \;{\rm km}\,{\rm s}^{-1}$.
For both of these modes the ejection is bipolar and parallel to the angular momentum vector of the SMBH accretion disc.
X-ray feedback from SMBHs is also implemented, but it has a minimal effect on the galaxy mass function.

All simulations from both suites follow the evolution of $2\times256^3$ dark matter plus fluid elements in a periodic comoving volume of $(25~h^{-1}{\rm Mpc})^3$. All simulations share the value of the following cosmological parameters: $\Omega_{\rm b}=0.049$, $h=0.6711$, $n_s=0.9624$, $\sum m_\nu=0.0$ eV, $w=-1$. However, each simulation has a different value of $\Omega_{\rm m}$ and $\sigma_8$. The simulations also vary the values of four astrophysical parameters that control the efficiency of supernova and SMBH feedback: $A_{\rm SN1}$, $A_{\rm SN2}$, $A_{\rm AGN1}$, and $A_{\rm AGN2}$. Details about the effect of these parameters and the range of values they can take is given in Table \ref{table:camels_params}.

The simulations with the different parameters are arranged into 4 sets.
In the LH set, which contains 1000 simulations for each code, values are arranged on a latin-hypercube, and each simulation has a different random seed for initial conditions.
We note that the latin-hypercubes of the IllustrisTNG and Simba simulations are different, i.e. there is no correspondence between simulations among the two sets.
The 1P set contains simulations in which a single parameter varies, and all other parameters are kept fixed at their fiducial value. 
There are 11 simulations for each parameter, meaning there are 61 simulations for each code.
The same initial conditions random seed is used for all the 1P simulations.
The CAMELS project also contains a cosmic variance and extreme set, but we do not use either of these in our work.

The merger trees available in the CAMELS project are created using the \textsc{ConsistentTrees} code \citep{consistent_trees} which is built on top of halos located using the \textsc{Rockstar} algorithm \citep{rockstar}. However the \textsc{Rockstar} halo catalogs do not contain all the galaxy properties we require for this work, such as star formation rate (SFR). Therefore we match the \textsc{Rockstar} and \textsc{SubFind} halos so that we have access to the data that we require. We use the method described in \cite{matching_halos}, but allow halos to be within 3x half mass radius to increase our sample size. Details about how well the catalogs are matched can be found in Appendix \ref{sec:camels_matching}.

\subsection{Input and output features}
\begin{figure}
	\includegraphics[width=\columnwidth]{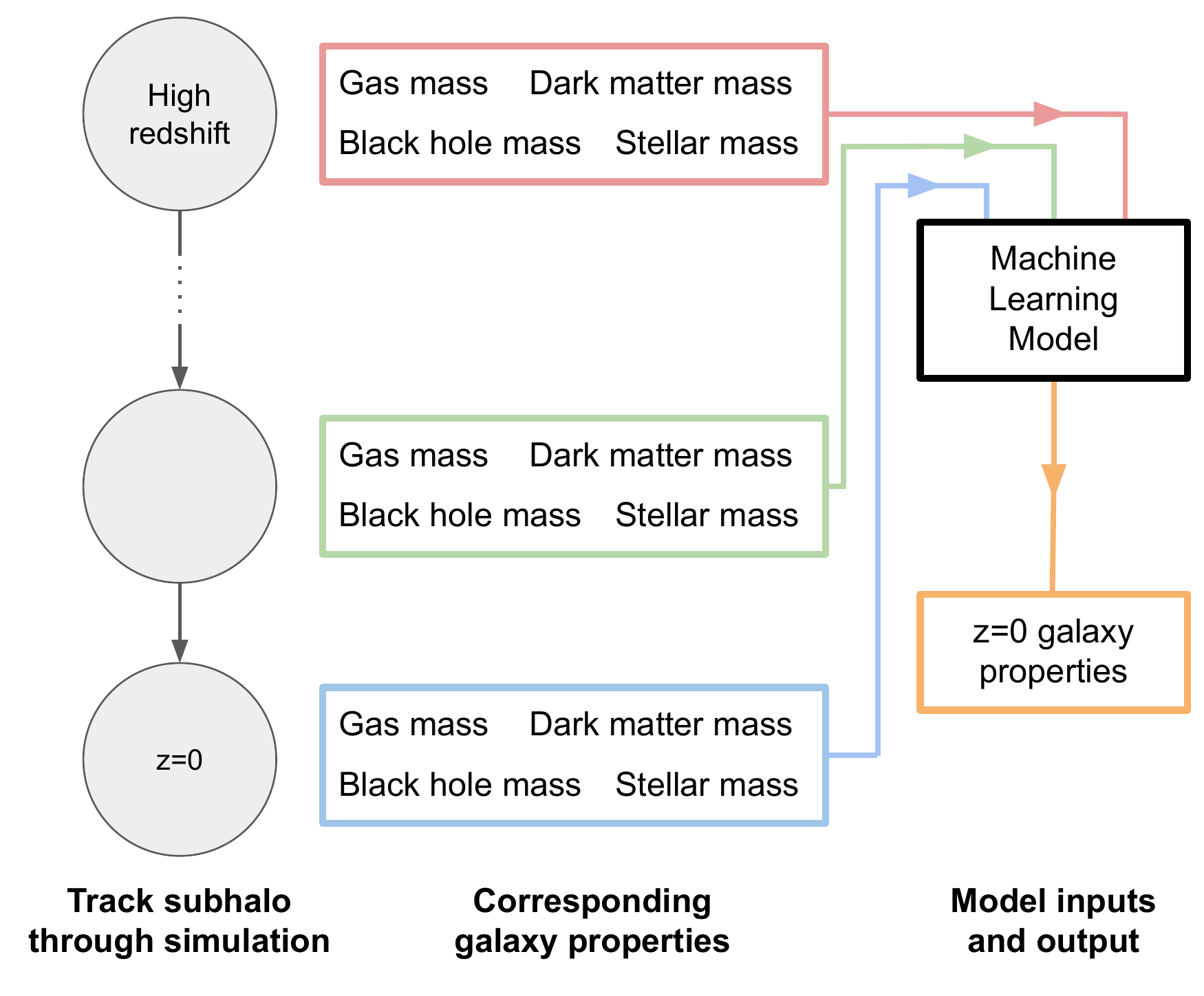}
    \caption{Summary of the method used in this work. The machine learning model takes in the four input features of the base model, but from a range of snapshots, not just redshift zero. The output for all models is the subhalo's baryonic properties at redshift zero.}
    \label{fig:method_summary}
\end{figure}

As in \cite{multi_epoch_1}, the inputs to our model are properties of a subhalo taken from a wide range of redshifts.
As decision trees are invariant to the scaling of the input features, therefore we do not scale the input features in any way, despite the fact that their values span multiple orders of magnitude.
As discussed in \cite{multi_epoch_1} we log the output features to prevent high mass galaxies being given a significantly higher weight than low mass ones.

\begin{figure}
	\includegraphics[width=\columnwidth]{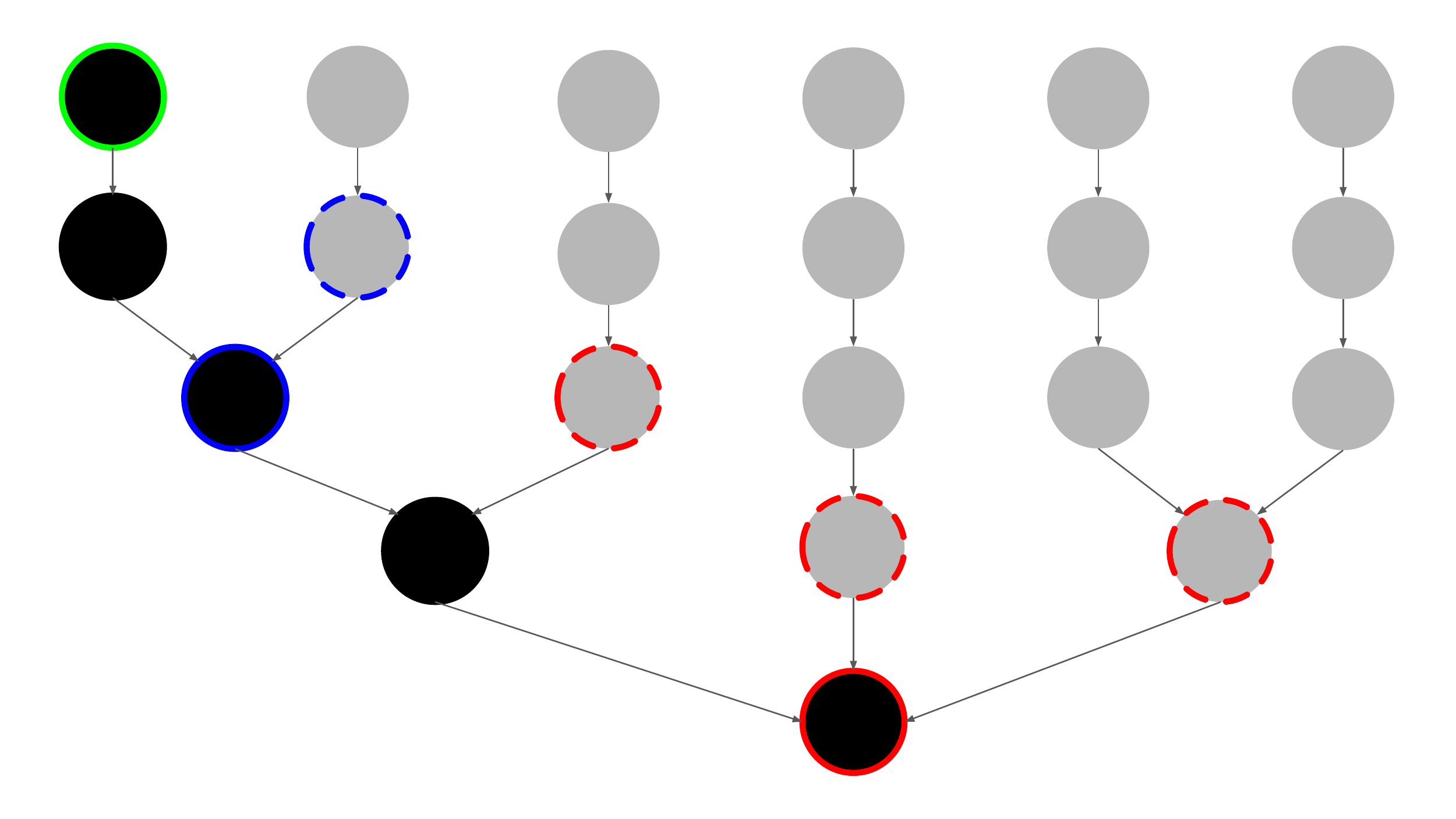}
    \caption{How subhalos that have merged are passed as input to the model. The black circles indicate the main progenitor branch of the merger tree, the grey circles indicates subhalos that merge with the main subhalo. Three snapshots are considered as input here, the green circle, blue circle, and red circle. The merger input feature is defined as the sum of the properties of the subhalos that have merged since the last snapshot that was used as input. This means that for the blue snapshot the single subhalo with the blue dashed line is used as the merger feature. For the red snapshot the merger feature is equal to the sum of the masses of the three subhalos with a red dashed outline. }
    \label{fig:merger_features}
\end{figure}
 
We do not consider every snapshot from the simulations as model input since this result in too large correlations in our input features for the feature importance to work effectively.
Therefore we use properties from every $d^{th}$ snapshot as model input. We choose $d$ for each simulation such that we get 10 snapshots approximately evenly spaced in time.
For each of these input snapshots we use the value of a number of the galaxy properties as an input, where $i$ denotes the property (gas mass, dm mass, bh mass, or stellar mass).
A summary of this part of our method is shown in Figure \ref{fig:method_summary}.

In \cite{multi_epoch_1} we only considered the main progenitor branch as input to our model. In this work we consider the impact of other branches that merge.
We define the merger feature at the $s^{th}$ snapshot for the $i^{th}$ property as
\begin{equation}
    M_s^i = \sum_{t=s-d}^{t=s} m_t^i
\end{equation}
where $m_t^i$ is the amount of mass of property $i$ that merged into the main progenitor branch at snapshot $t$. 
Thus we are summing the mass of all the subhalos that merge into the main progenitor branch between the input snapshots.
This method is not able to distinguish between a large number of minor mergers and a single major merger, but it still captures information about the importance of discrete vs smooth accretion for the halo's evolution.
A schematic of this process is shown in Figure \ref{fig:merger_features}.

In this work we show the results from predicting four different output features,  although this method could be used to gain information about any galaxy property. We predict the $z=0$ stellar and gas mass, which are given by the total mass of all stellar/gas particles/cells identified by \textsc{SubFind} as bound to the subhalo. The galaxy SFR is defined as the sum of the individual star formation rates of all gas elements in the subhalo. The stellar metallicity is given by the mass-weighted average metallicity of the star particles.

\subsection{Machine learning algorithms}

\subsubsection{ERT}
Random forest regressors \citep{rf_1,rf_2} use an ensemble of decision trees to make a prediction. Each decision tree is constructed top-down from a root node. At each node the data is split into two bins based on the values of its input parameters. The splits are chosen such that the weighted average of the mean squared error (MSE) of the two bins is minimized \citep{rf_3}. This partitioning of the data results in each leaf node at the bottom of the tree containing a small subset of the data, where almost all members of the subset have a similar output value. Predictions from decision trees are based on the assumption that test data points will have a similar output value to the other members of the leaf node it is placed into. A random forest is made up of a number of decision trees. There is a bootstrapping procedure such that each decision tree within the forest is trained on a randomly generated subset of the training data. Further randomness is added in that for each split only a subset of input features can be used. The prediction from a random forest is the average prediction of its component decision trees. A major advantage of random forests is that they are significantly less prone to overfitting data compared with a single decision tree. This results from the randomness added when training the individual decision trees.

For this work we use extremely randomised tree ensembles (ERT; \cite{ert}). This is the algorithm used in previous work \citep[e.g.][]{kamdar_painting, dave_painting_1, mssm_painting, dave_painting_2, lovell_painting, multi_epoch_1} for predicting galaxy properties, and we found it to slightly outperform the standard random forest. It adds in additional randomization by computing a random split for each feature at each node, rather than the optimal split.

\begin{figure}
	\includegraphics[width=\columnwidth]{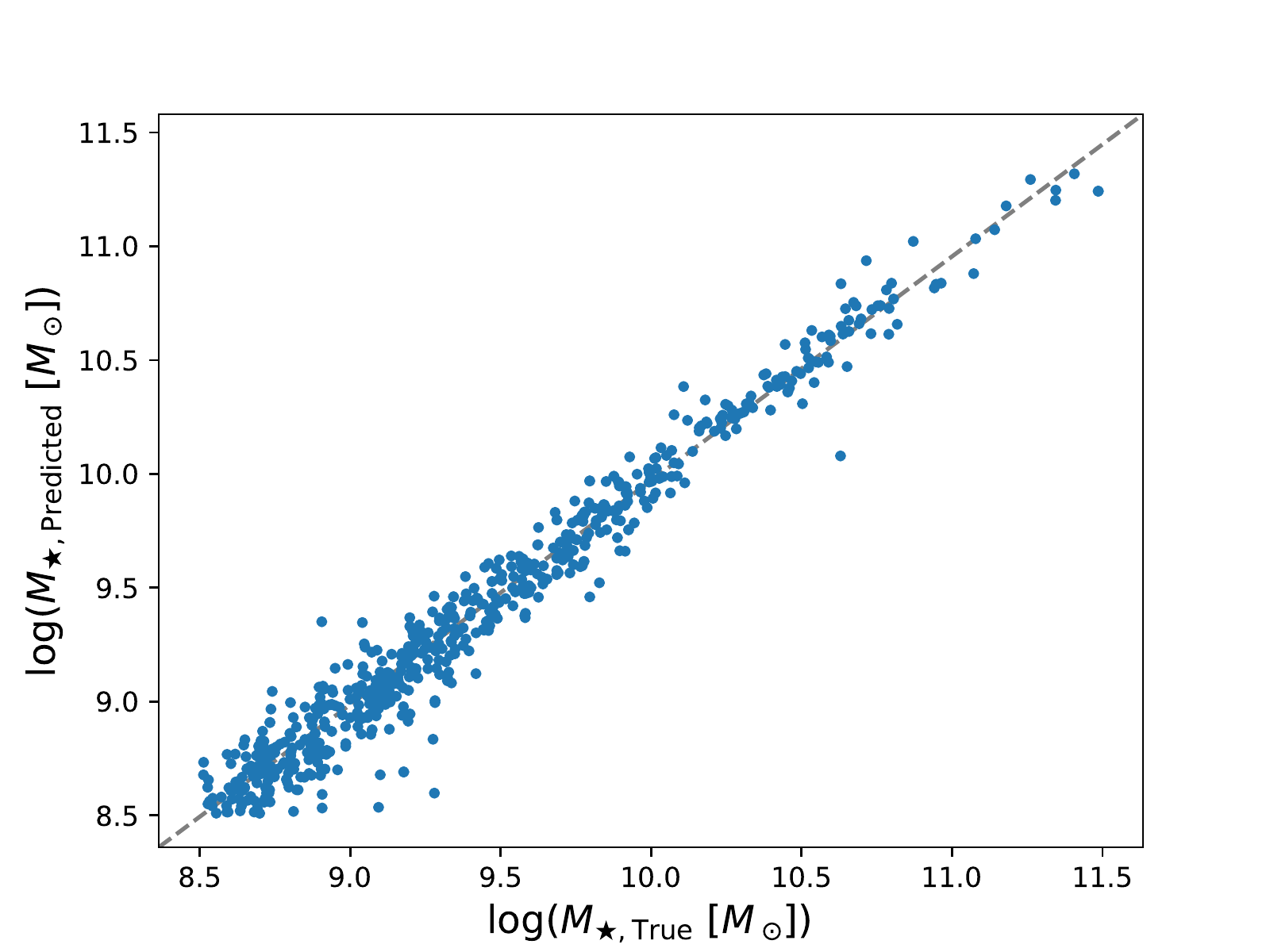}
    \caption{Model performance on 1000 randomly sampled galaxies from the TNG100 test set. The x axis shows the stellar mass from the simulation, and the y axis shows the value predicted by the machine learning model. This shows that the model has learned the relationship between the input features and stellar mass.}
    \label{fig:model_performance}
\end{figure}

One major benefit to using ensembles based on decision trees is the ability to extract information on which input features are providing the information that is used to make the final predictions. For each decision tree the importance of an input feature can be determined by the number of times it is used for a split, and how close to the top of the tree those splits are. By averaging the importance values over all the decision trees within the ensemble model, a set of the feature importances can be determined for the model as a whole. However, one must be aware that correlations between input features will affect their importance values, and can make the results more difficult to interpret. 
When examining feature importance plots the differences in the relative importance of each input feature should be considered, rather than their absolute values.

\subsubsection{Principal Component Analysis (PCA)}
In order to help visualize the results of applying our method to the CAMELS simulations, we make use of PCA \citep{pca_1, pca_2} to reduce the dimensionality of the data. PCA works by determining a new coordinate system, where each axis is a linear combination of the the original data axes. The first axis (known as the first principal component) represents the direction of maximum variance, and subsequent axes represent directions of decreasing variance.
We choose to use PCA over other available non-linear dimensionality reduction methods as it allows us to easily extract information about the reduced components that the algorithm finds.
We verify that the dimensionality reduction is not significantly different when using the UMAP algorithm \citep{umap} instead.

\section{Subsamples in a simulation}
\label{sec:subsamples}

\subsection{Is the model learning relationships?}
In this work we show how the feature importance changes for different simulations and different subsamples of galaxies. In order for the feature importance to be meaningful, we need to ensure that the model has been able to successfully learn a relationship between the input and output features. In Figure \ref{fig:model_performance} we show the true vs predicted stellar mass value for 1000 randomly sampled galaxies from TNG100. A model that made perfect predictions would correspond to all points lying on the diagonal. The small scatter in the figure tells us that the ERT model has successfully learnt a function mapping the input features to stellar mass, so the feature importance values can be trusted. 
The MSE scores for all the models used in this work are shown in Appendix \ref{sec:model_performance}.

\subsection{Reading feature importance plots}
\label{subsec:reading_fi_plots}

\begin{figure}
	\includegraphics[width=\columnwidth]{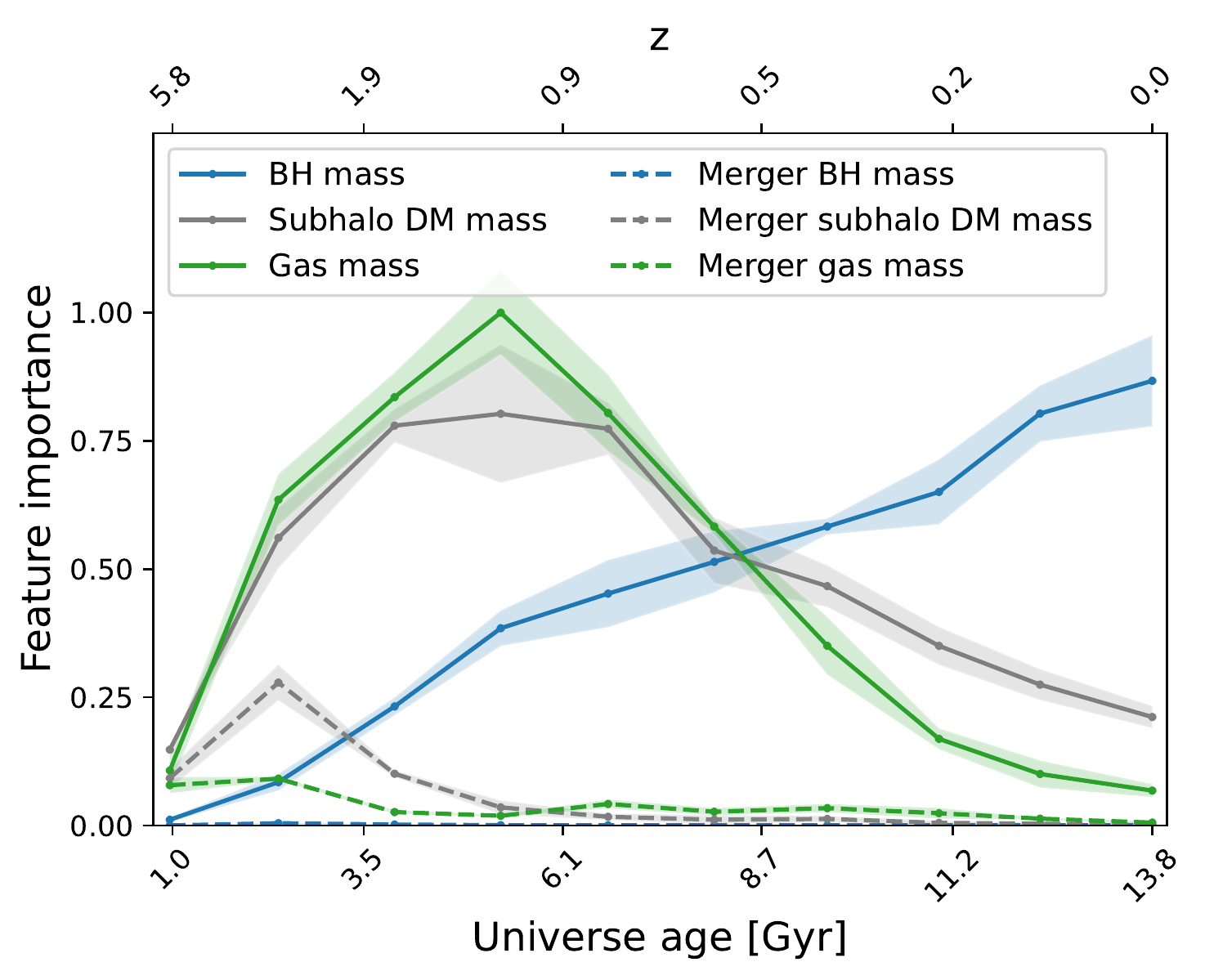}
    \caption{Feature importance of an ERT model trained to predict stellar mass of galaxies in the TNG100-1 simulation. Merger trees were generated using the LHaloTree algorithm.}
    \label{fig:fi_tng}
\end{figure}

\begin{figure*}
	\includegraphics[width=\textwidth]{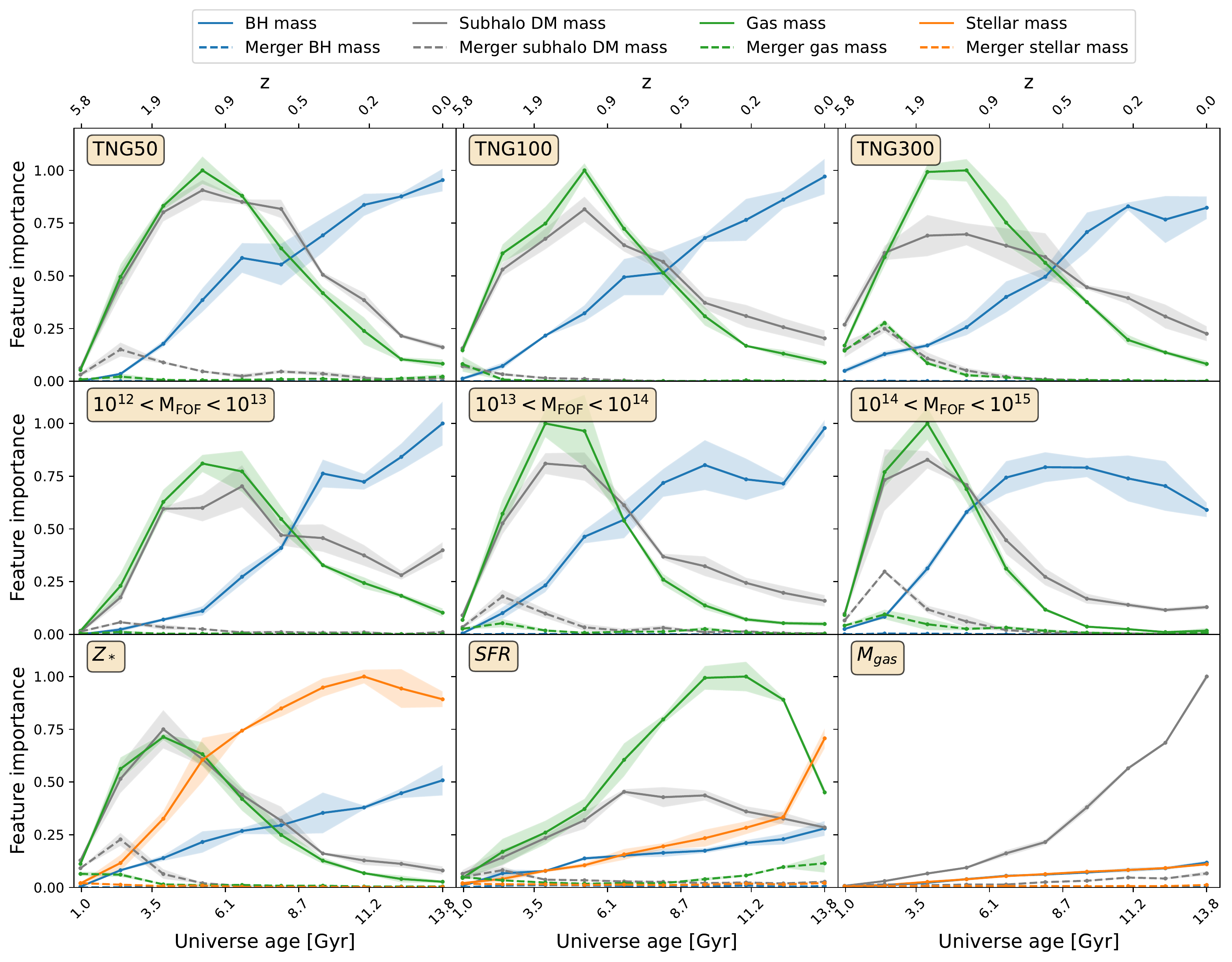}
    \caption{Feature importance plots from models trained on IllustrisTNG, so all panels have the same subgrid models. Models are trained to predict $z=0$ stellar mass unless otherwise indicated. The legend at the top of the figure is shared across all panels. \textbf{Top left} TNG50 - higher resolution simulation, \textbf{Top centre} TNG100 - \textsc{SubLink} merger trees, \textbf{Top right} TNG300 - lower resolution simulation, \textbf{Middle left} $10^{12} < \mathrm{M_{FOF}} < 10^{13}$ - Low density environment, \textbf{Middle centre} $10^{13} < \mathrm{M_{FOF}} < 10^{14}$ - Medium density environment, \textbf{Middle right} $10^{14} < \mathrm{M_{FOF}} < 10^{15}$ - High density environment, \textbf{Bottom left} Predicting $z=0$ stellar metallicity, \textbf{Bottom centre} Predicting $z=0$ SFR, \textbf{Bottom right} Predicting $z=0$ gas mass}
    \label{fig:fi_subsamples}
\end{figure*}

Figure \ref{fig:fi_tng} show the feature importance obtained from a model trained to predict $z=0$ stellar mass of galaxies in the TNG100 simulation.
Merger trees were extracted using the \textsc{LHaloTree} algorithm.
Each point on the plot corresponds to the importance of an input property at a certain time in the simulation.
We include all input properties other than the one we are predicting, e.g. we do not use stellar mass as an input feature when predicting stellar mass.
The maximum value of the feature importance is normalized to one for all models, so we only consider the relative importance of the input properties at different times rather than focusing on the absolute values.
We highlight the fact that a large feature importance value does not necessarily mean that the input feature has a large value at that point.
For example, in Figure \ref{fig:fi_tng} the feature importance for the dark matter mass peaks at early times then drops off.
This does not mean that the halos within IllustrisTNG are decreasing in dark matter mass, instead it indicates that the dark matter mass at early times is more informative for predicting the stellar mass at $z=0$ than the dark matter mass at late times.
We also note that a high feature importance value does not mean that the input feature has a positive correlation with the output feature being predicted.
For example, we find that black hole mass is an important factor when predicting SFR, but for large galaxies black hole mass is negatively correlated with SFR.
To give an error on the feature importance we train a model for 10 different train/test splits of the data.
The shaded region in Figure \ref{fig:fi_tng} corresponds to the standard error taken on the feature importance values from the 10 different models.

We now interpret the feature importance plot shown in Figure \ref{fig:fi_tng}.
We see that the dark matter and gas mass feature importance peak at early times, then decrease. 
Physically this corresponds to the initial period of formation when the universe SFRD is highest and most stars are being formed.
The black hole feature importance is similar in magnitude to that from gas and dark matter mass, but peaks at late times.
Relative to the importance of the main progenitor branch, the feature importance of mergers is very small, and peaks earlier than the main progenitor feature importance.
Black holes from mergers are deemed to be completely unimportant.
This is unsurprising as most halos that merge will be too small to host a black hole.
However, the fact that the feature importance for black hole mergers is zero provides evidence that our model is not overfitting, since an overfitted model would end up using uninformative features to make splits.

\subsection{Comparing feature importance plots}

Figure \ref{fig:fi_subsamples} contains a number of feature importance plots from models trained on the IllustrisTNG simulation suite.
In the top row we show the effect of varying resolution.
The top centre plot gives the feature importance of a model trained to predict the stellar mass of TNG100 galaxies. This is the same as Figure \ref{fig:fi_tng}, except the merger trees were generated using the \textsc{SubLink} algorithm.
We see the same trends in the two plots, with all progenitor input properties peaking at the same point, and having the same relative importance.
There is a minor difference in the importance of the merger features, with them being deemed less important for the \textsc{SubLink} merger trees.
However, the overall agreement provides evidence of the robustness of this method to the choice of merger tree algorithm.
For the remaining IllustrisTNG plots we use the \textsc{LHaloTree} merger trees.

\begin{figure*}
	\includegraphics[width=\textwidth]{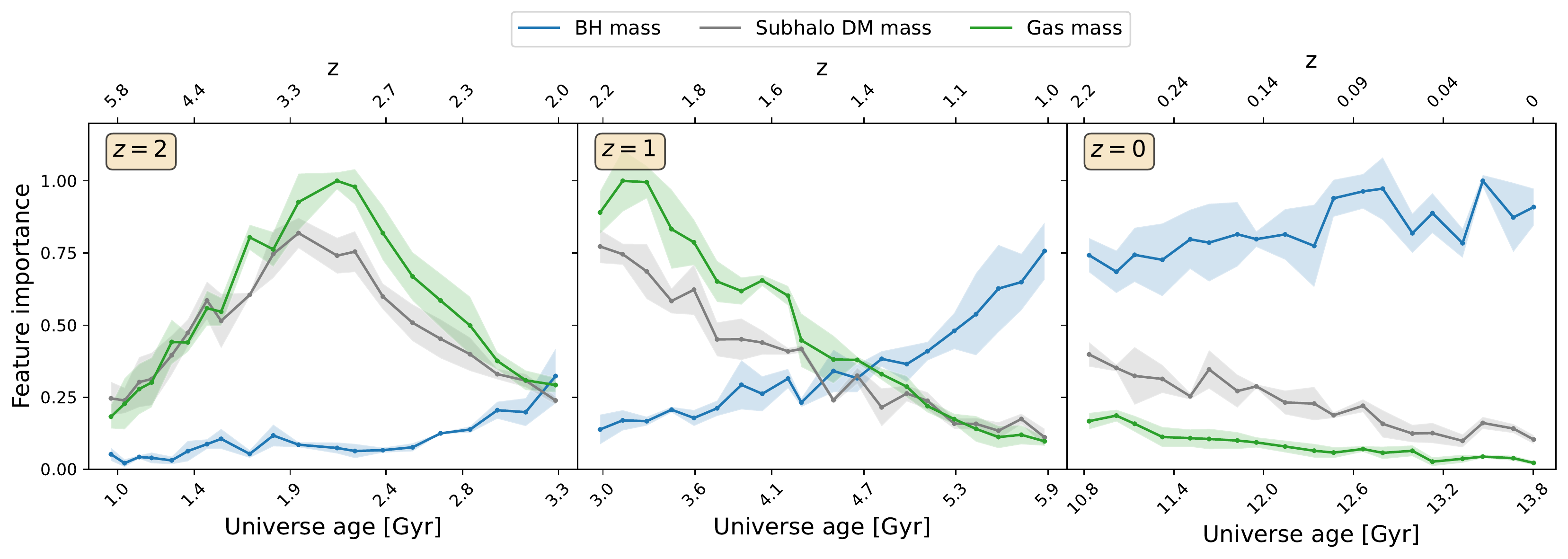}
    \caption{Feature importance plots for predicting the stellar mass of TNG100 galaxies at different redshifts. \textbf{Left} $z=2$, \textbf{Centre} $z=1$, \textbf{Right} $z=0$. The legend at the top of the figure is shared across all panels.}
    \label{fig:fi_different_z}
\end{figure*}

The top left and top right panels show a model trained using galaxies from TNG50 and TNG300 respectively.
The general trends are very similar to the model trained on TNG100, but some differences appear.
For all three resolutions the BH mass peak has a similar value as the gas mass peak.
For TNG50 the peak of the gas mass feature importance has a similar magnitude to the peak of the dark matter mass, whereas for TNG300 the peaks are further apart, and for TNG100 the distance between peaks is intermediate.
Thus there is a clear trend in decreasing dark matter feature importance with decreasing resolution.
This trend also occurs in the feature importance plots from models trained on the lower resolution Illustris simulations.
In low resolution simulations the deep potentials at the centre of halos cannot be fully resolved. This means they have less ability to hold on to baryons since stellar feedback is capable of driving gas further from the ISM, significantly impacting star formation.
Thus this method allows us to quantify the impact of resolution as we can clearly see at what resolution the feature importance plots start to diverge.
There are no significant differences in the black hole feature importance.
The peak of star formation appears to occur at the same point for all three resolutions. In \cite{2020MNRAS.493.2926L} a suite of simulations was run using the EAGLE model with fixed particle mass, but the force softening scale was varied. They found that the cosmic star formation history becomes increasingly biased toward high-redshift, but that the effect was small for the range of values used by the different TNG simulations.

In the middle row of Figure \ref{fig:fi_subsamples} we show the feature importance from models trained on galaxies taken from different density environments. 
All galaxies are taken from TNG100 and are split into three samples based on the mass of their FOF halo.
The bin edges are given by $\log{\mathrm{M_{FOF}}} = 12,13,14,15$.
The time at which the peak in gas and dark matter importance occurs shifts as the environment is varied.
For galaxies in low density areas the peak occurs at later times, indicating delayed galaxy formation.
This agrees with the findings of \cite{2022ApJ...941....5J} who examined the star formation history (SFH) of a range of galaxies from the Horizon-AGN simulation \citep{horizon_agn_1}, showing that IllustrisTNG and Horizon-AGN are in agreement in this area. These results are also consistent with observational studies which determine SFHs using stellar population modelling \citep{2005ApJ...621..673T, 2015MNRAS.450.2749G}.
For the galaxies in high density regions the majority of star formation occurs at earlier times.
This also causes the black hole feature importance to peak prior to $z=0$.
We find that the $z=0$ black hole feature importance decreases with increasing density, in agreement with the observational results of \cite{2022MNRAS.509.1805C}.
As we increase density the importance of merger features relative to the progenitor features increases. This is to be expected as galaxies within groups and clusters will experience a large number of mergers, albeit at early times.

In the final row of Figure \ref{fig:fi_subsamples} we show the feature importance from models trained to predict other properties of TNG100 galaxies at $z=0$.
The left, centre, and right panels correspond to output features of stellar metallicity, SFR, and gas mass respectively.
The feature importance of dark matter and gas, and black hole mass in the stellar metallicity plot is similar to that from stellar mass plots.

The feature importance plot for SFR is significantly different to those for stellar mass, with features peaking close to or at $z=0$.
This is because SFR is an instantaneous property unlike stellar mass which builds up over time.
As the dark matter mass gives the gravitational potential, which indicates how much gas will fall onto the halo. However, gas must cool before it can form stars, which explains why the dark matter feature importance peaks at earlier times than any other input property, as any gas which was recently accreted is unlikely to contribute to star formation.
Of the merger features only gas mass is important, but it does indicate that mergers have a minor effect on the overall galaxy population SFR at $z=0$.

For the prediction of gas mass the dark matter mass dominates.
This is because we are predicting the gas mass of the halo, of which the majority is hot gas, rather than the mass of the ISM.
Stellar mass and black hole mass do have some importance, but this plot shows that in general the feedback in the IllustrisTNG model is insufficient to eject gas from halos.

\subsection{Predictions at different redshifts}

Figure \ref{fig:fi_different_z} shows models trained to predict stellar mass for galaxies from TNG100 at different redshifts.
We wish to consider how the relative importance of the various input properties changes when considering predictions of stellar mass at different times. 
The left, centre, and right panels show predictions for $z=2,1,0$ galaxies respectively.
Due to the age of the universe at $z=2$ we can only use galaxy properties from the past 3Gyr as input features. We therefore also restrict the inputs to the $z=1$ and $z=0$ models to the past 3Gyr to allow for a direct comparison.

The leftmost $z=2$ shows a clear peak in gas and dark matter mass around $z=3$.
This is despite the fact that there is still a large amount of star formation ongoing at $z=2$.
However, the gas needs time to collapse into the gravitational well of the halo, as well as radiate away energy, before it is able to form stars.
This explains why the peak is located about 1Gyr prior to $z=2$.
In the $z=1$ panel we see gas and dark matter mass continuously dropping, indicating that 3Gyr before $z=1$ is already past the peak of star formation.
However, the black hole feature importance continues to increase, indicating it's coupling to the stellar mass of galaxies.
For the $z=0$ plot the feature importance is nearly flat across time.
This is a reflection of the lack of star formation in the majority of galaxies in the present epoch.
A trend across the three panels is the increasing importance of black holes.
This is because it takes time for black holes to build up and become effective, at $z=2$ they are not that massive for most galaxies.
The right panel of Figure \ref{fig:fi_different_z} can be compared with the last 3 Gyr of Figure \ref{fig:fi_tng}. When we look at the feature importance in this range of Figure \ref{fig:fi_tng} we see BH mass is the most important feature, and its importance is increasing with time. We see that the importance of gas mass and DM mass is decreasing with time. These trends are in agreement with Figure \ref{fig:fi_different_z}.

\section{Comparing simulations}
\label{sec:compare_sim}

\begin{figure*}
	\includegraphics[width=\textwidth]{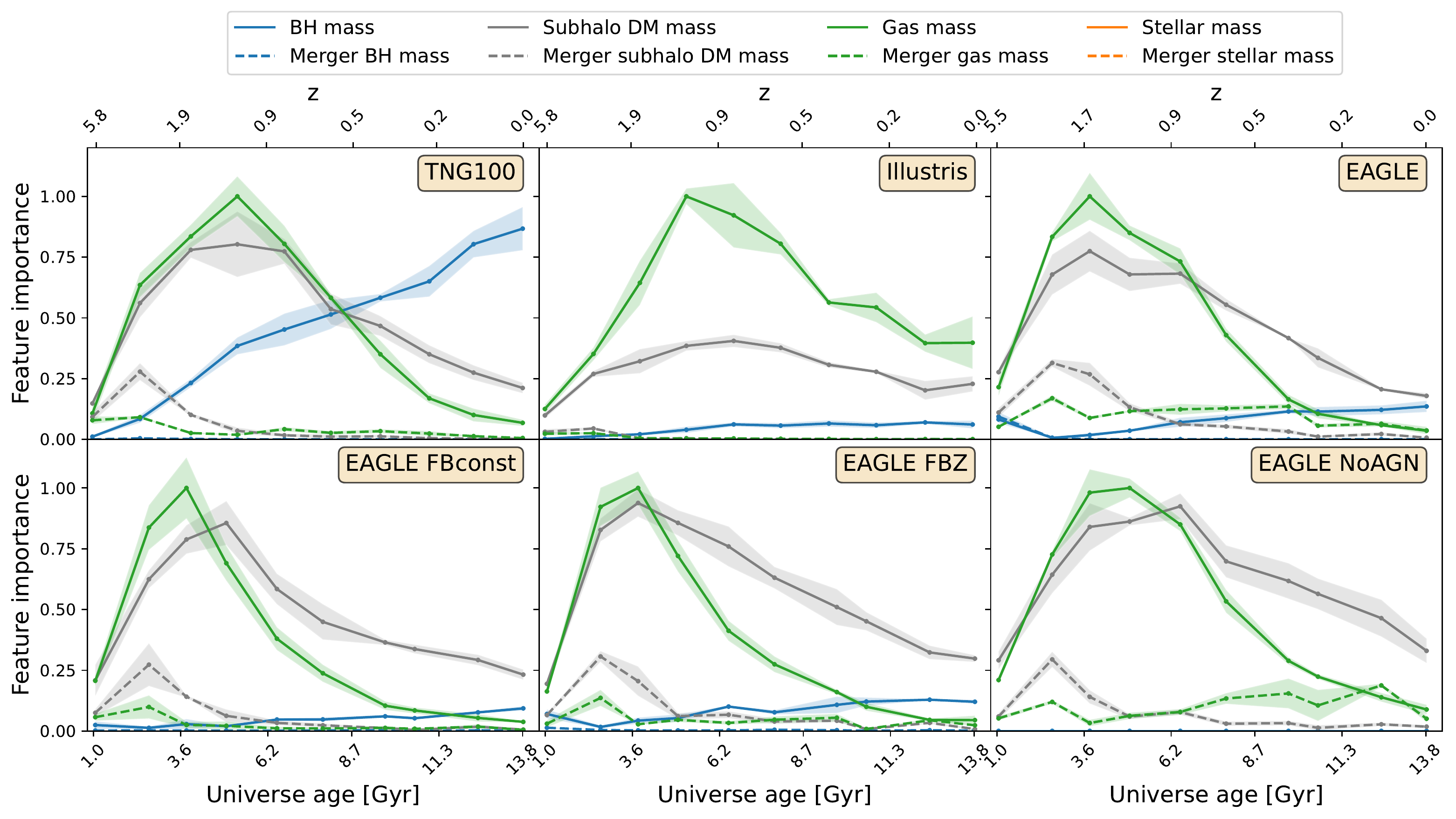}
    \caption{Feature importance plots for predicting stellar mass of simulations with different subgrid models. The legend at the top of the figure is shared across all panels. \textbf{Top left} TNG100, \textbf{Top centre} Illustris, \textbf{Top right} EAGLE fiducial, \textbf{Bottom left} EAGLE FBconst, \textbf{Bottom centre} EAGLE FBZ, \textbf{Bottom right} EAGLE NoAGN. As the simulations have slightly different cosmologies, the horizontal axis are not exactly aligned.}
    \label{fig:fi_different_sims}
\end{figure*}

In Figure \ref{fig:fi_different_sims} we compare the feature importance from models trained on simulations with different subgrid model implementations.
As these simulations have slightly differing cosmologies, the horizontal axes are not exactly aligned, but they are close enough for comparisons to still be valid.
Figure \ref{fig:stellar_mass_function} contains the stellar mass function for TNG, Illustris, and EAGLE. Within the stellar mass range of the training data there is reasonable agreement between the simulations. Restricting the mass range further still yields different feature importance plots between simulations.
The top left panel shows the results from TNG100, as discussed in section \ref{sec:subsamples}. 
The top centre panel shows a model trained on the original Illustris simulation. 
There are three differences when compared with the TNG. 
Firstly the gas mass importance relative to dark matter is significantly increased for Illustris.
Secondly the peaks in gas and dark matter mass are less pronounced for Illustris, and they occur at later times compared to the TNG. 
Both of these differences are reflections of the changes to the supernova feedback implementations. 
The feedback in TNG is more effective, which means a deep gravitational potential is needed to hold on to gas in star forming halos.
This explains why dark matter importance increases for the TNG.
The reduced efficiency of stellar feedback in Illustris means that star formation can continue for longer, which is reflected in the shape and location of the peaks.
The third difference is the relative importance of black holes, which are much more prominent in the TNG.
This is a result of the new black hole feedback mode introduced in the TNG which boosts black hole feedback at low accretion rates.
The merger trees for the Illustris simulation are created using the \textsc{SubLink} algorithm, which explains the differences in the merger feature importance.

The top right panel shows the results of the model trained on the fiducial EAGLE simulation.
The gas mass and dark matter mass are close to those from the TNG, but the black hole importance is similar to Illustris.
This is interesting as EAGLE is run with an SPH code, but both TNG and Illustris use a moving mesh to model the gas hydrodynamics.
The fact that the EAGLE gas mass and dark matter mass importances are much closer to the TNG than the TNG is to Illustris shows that it is the subgrid models that are to first order key to determining the correct build up of galaxy properties, rather than the hydrodynamics solver which is used.
This agrees with the results of \cite{aquila} who simulated an individual Milky Way-like galaxy using multiple cosmological hydrodynamical codes.
The black hole importance shows that AGN feedback in EAGLE has similar efficiency to that in Illustris, despite the significantly different implementations.

The bottom right panel shows the EAGLE run without any black holes.
Compared with the fiducial EAGLE run the gas and dark matter mass is more important at late times in the NoAGN run.
This confirms that AGN do have any effect in shutting off some star formation in EAGLE.

The bottom left panel shows the EAGLE run where supernova feedback is independent of environment.
The ISM dependence in the fiducial run makes feedback more efficient for low mass galaxies.
Thus when the feedback is constant more star formation can occur in low mass galaxies, which means more stellar mass will build up at early times.
This is reflected in the fact that the gas and dark matter peaks move to the left for the FBconst plot.
There is also a decrease in the black hole importance.

The bottom centre panel displays the EAGLE run where supernova feedback depends only on metallicity, unlike the fiducial run where it also depends on density.
The peak occurs at a similar time to the FBconst run, showing that it is the density rather than the metallicity which is the factor determining the location of the peak of star formation density.
However, the dark matter feature importance is more similar to the fiducial run, indicating the metallicity dependence is responsible for this feature.

\section{Camels}
\label{sec:camels_results}
In this section we apply  the method to the CAMELS suite. We first focus on the effect of varying the $A_{SN2}$ parameter in the IllustrisTNG simulations from CAMELS, then compare the other parameters and the Simba simulations. 

\subsection{Correlations between supernova feedback and PCA components}
\begin{figure}
	\includegraphics[width=\columnwidth]{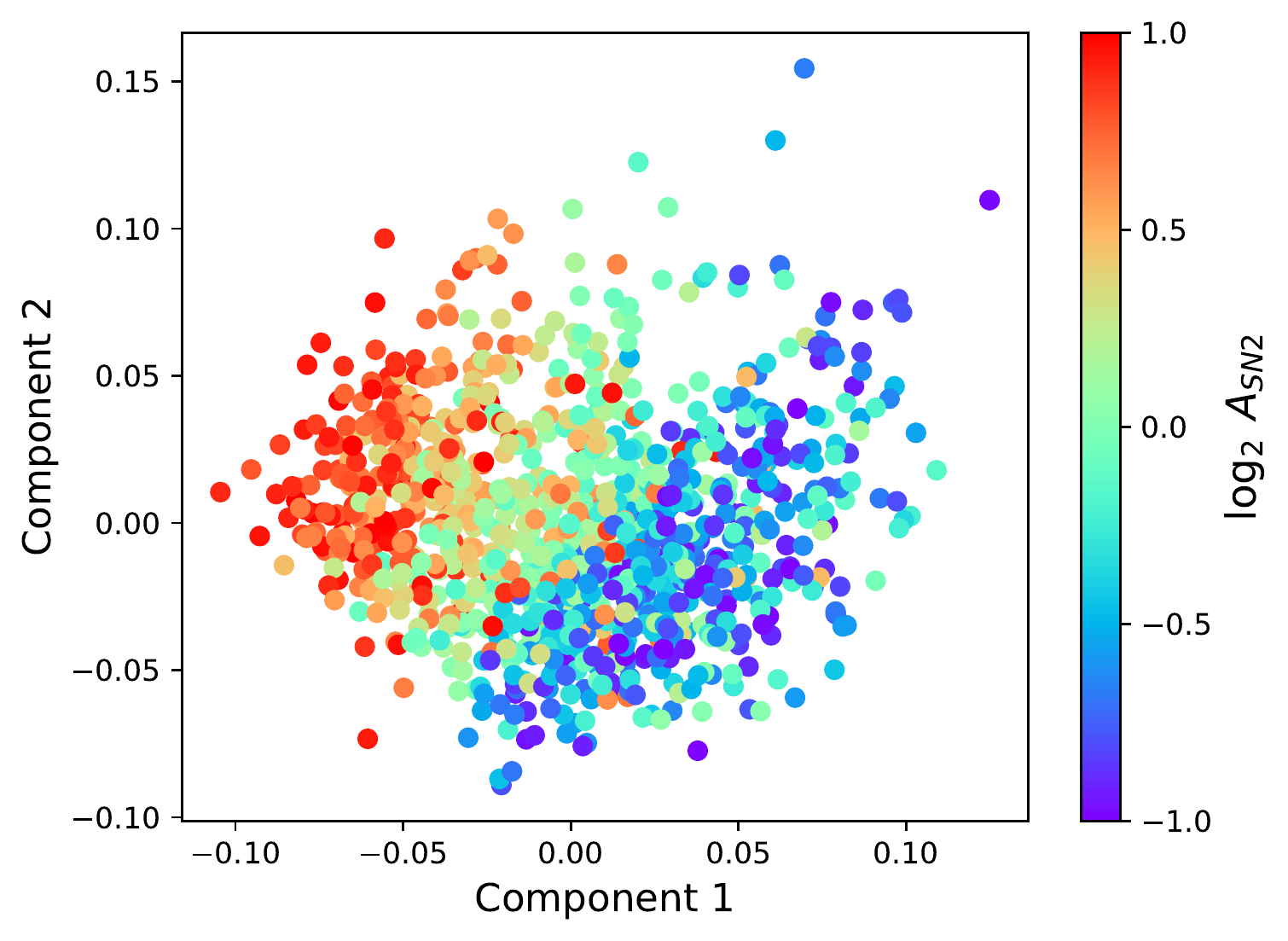}
    \caption{PCA plot of stellar mass feature importance applied to the IllustrisTNG CAMELS simulations. Each point represents a single simulation. Points are colored by the speed of the supernova winds within the simulation.}
    \label{fig:camels_pca}
\end{figure}

For each of the $N=1061$ IllustrisTNG simulations in the CAMELS suite we train an ERT model to predict the stellar mass at $z=0$.
From each of these models we extract the feature importances, and concatenate them into a vector.
The feature importance vector from each model has a length of $M=30$, which corresponds to 3 input properties (black hole mass, dark matter mass, gas mass) at 10 different snapshots.
Combining the feature importances from all the simulations gives us a matrix of size $N \times M$.
We apply PCA to this matrix, and show the results in Figure \ref{fig:camels_pca}.
In this plot each point corresponds to a single simulation.
The horizontal axis corresponds to the first PCA component, and the vertical axis corresponds to the second PCA component.
We color each of the points by the value of the $A_{SN2}$ parameter of the simulation.
Since $A_{SN2}$ is sampled uniformly in log space for the LH and 1P sets, the colorbar is also logged. 

\begin{figure*}
	\includegraphics[width=\textwidth]{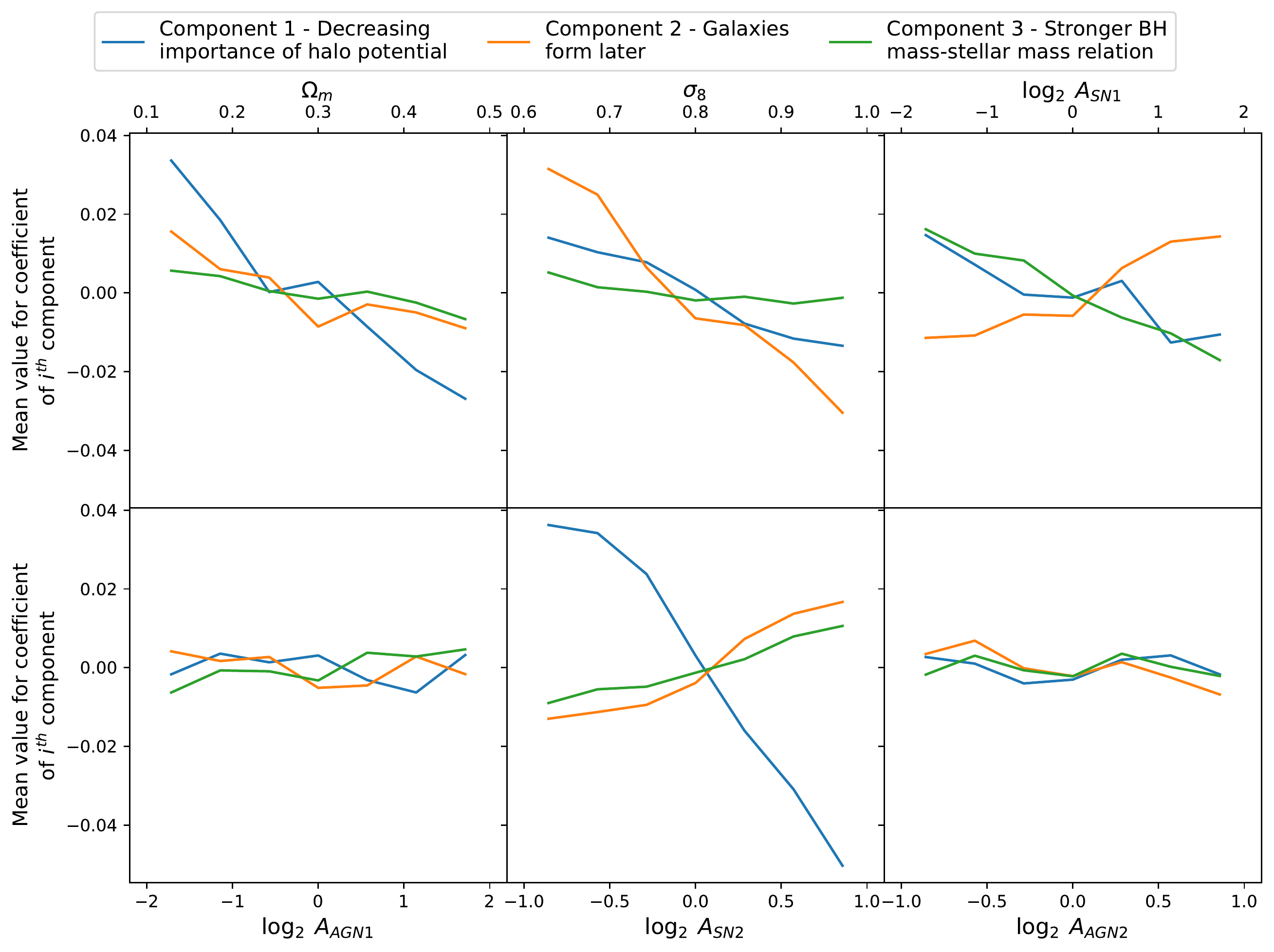}
    \caption{The effect of varying the simulation parameters on the PCA components. Models are trained on the IllustrisTNG simulations from CAMELS. \textbf{Top left} Omega matter, \textbf{Top centre} Sigma 8, \textbf{Top right} $A_{SN1}$, \textbf{Bottom left} $A_{AGN1}$, \textbf{Bottom centre} $A_{SN2}$, \textbf{Bottom right} $A_{AGN2}$. For a description of each parameter see Table \ref{table:camels_params}.}
    \label{fig:camels_tng_coefficients}
\end{figure*}

We can see that there is a clear trend with supernova feedback in the value of the PCA components, with a large $A_{SN2}$ value corresponding to a low component 1 coefficient.
There is a slight trend in component 2, but this is minor compared to the variation in component 1.
Rather than showing scatter plots for all the PCA components, we summarise the information in Figure \ref{fig:camels_pca} in the bottom centre panel of Figure \ref{fig:camels_tng_coefficients}.
Here we plot the mean value of the coefficient of the $i^{th}$ component for different $A_{SN2}$ bins.
This shows the negative correlation between the value of $A_{SN2}$ and the first PCA component, along with the minor trend in the second component.
This plot also allows us to examine other PCA components.
We can see that there is some correlation with $A_{SN2}$ and the third component.
We have not plotted the fourth component to avoid overcrowding, but it does not show any correlation.

\begin{figure}
	\includegraphics[width=\columnwidth]{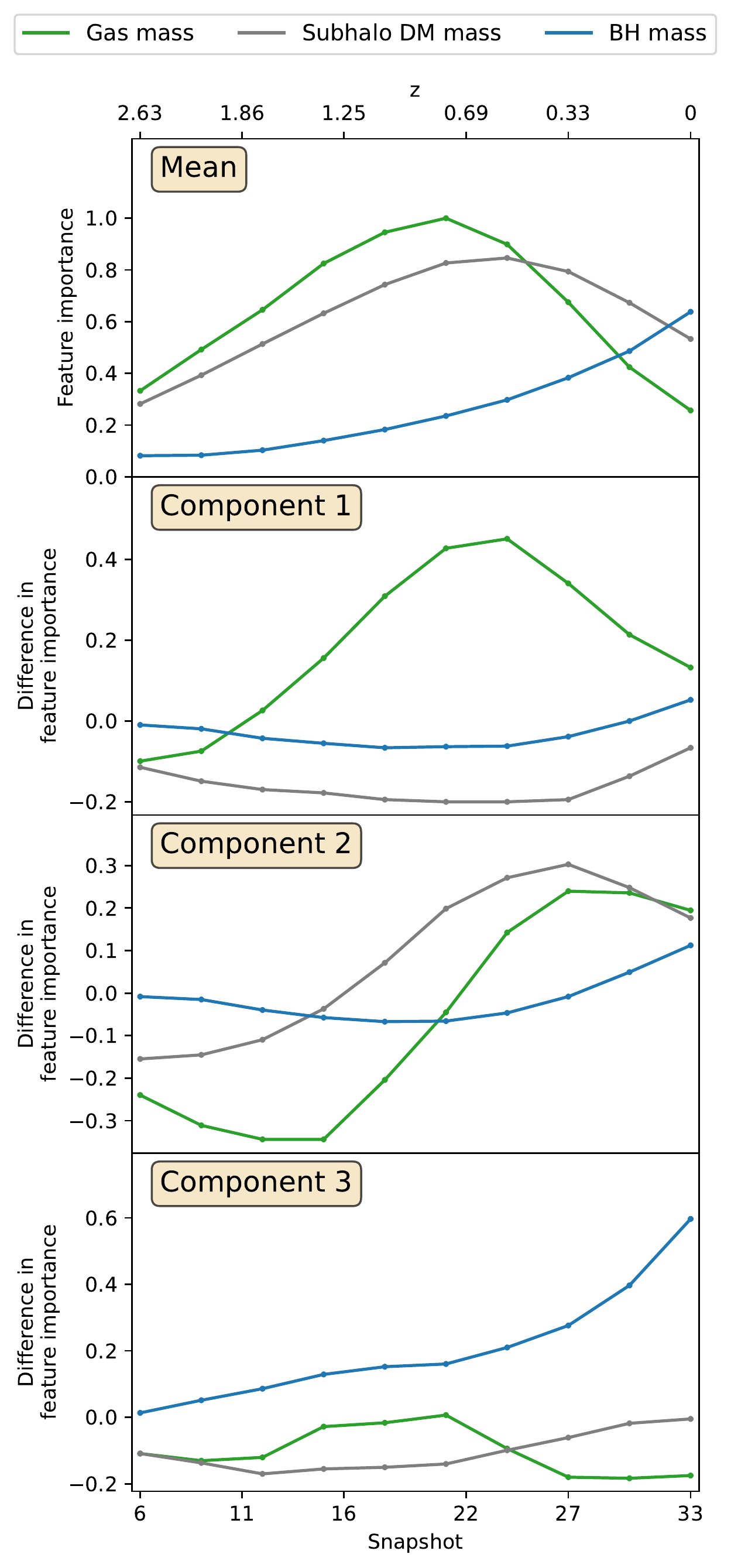}
    \caption{Plots of the value of each component resulting from PCA applied to the feature importance vectors of models trained to predict the stellar mass of galaxies from the CAMELS simulations. The legend at the top of the figure is shared across all panels.}
    \label{fig:camels_components}
\end{figure}

\begin{figure*}
	\includegraphics[width=\textwidth]{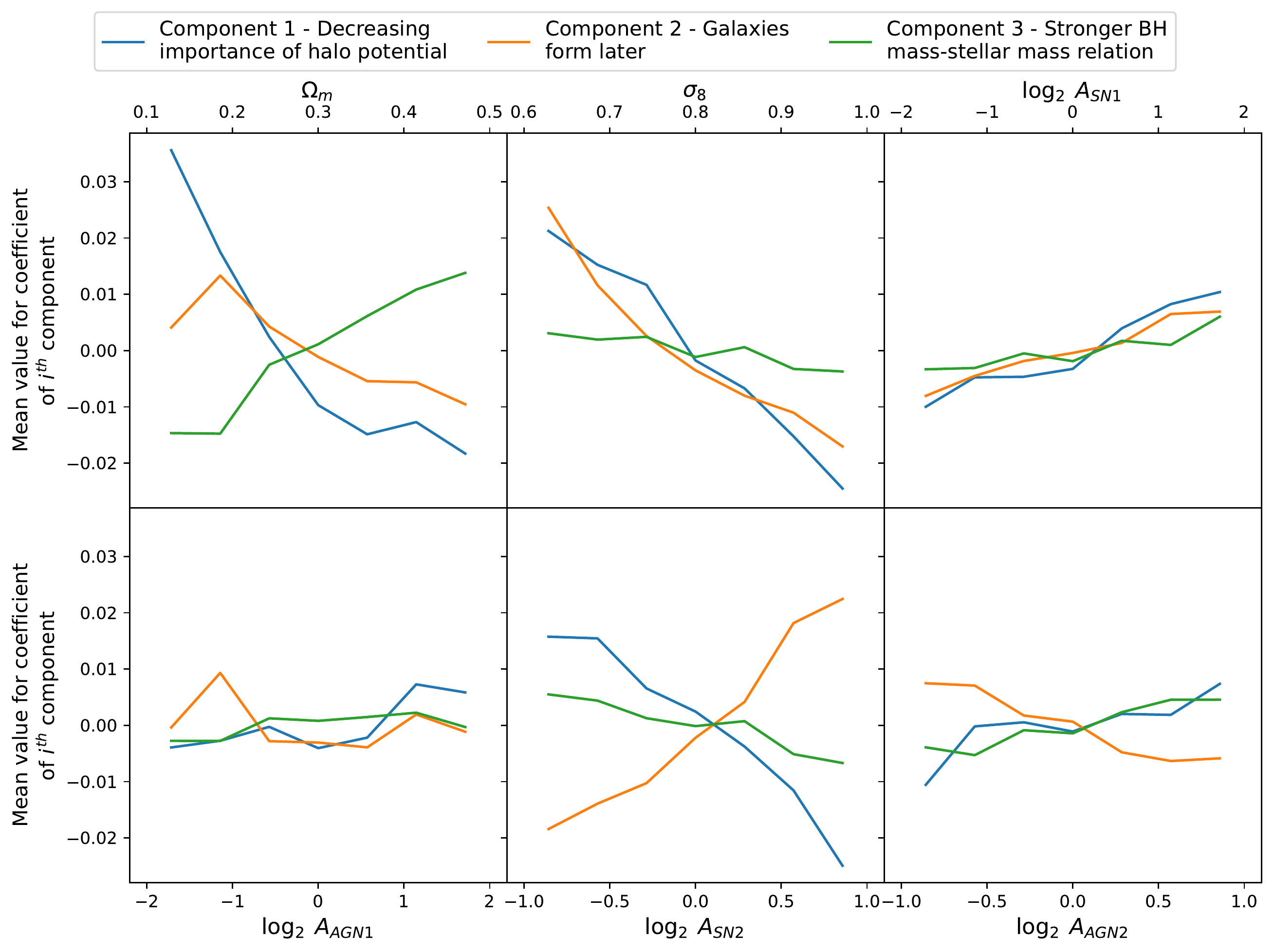}
    \caption{Same as Figure \ref{fig:camels_tng_coefficients}, but using the Simba simulation from the CAMELS suite. Components are determined from PCA applied to the IllustrisTNG simulations only}
    \label{fig:camels_simba_coefficients}
\end{figure*}

\subsection{Physical interpretation of PCA components}
Now that we have established the relationship between each of the PCA components and $A_{SN2}$ we wish to determine what each component corresponds to physically. To do this we plot the feature importance of the PCA mean and each of the first three components in Figure \ref{fig:camels_components}.
We would expect the feature importance of the mean to be similar to the one obtained from TNG100 (shown in Figure \ref{fig:fi_tng}).
We find the relative importance of each of the input properties to be similar, but the peak occurs later.
This is a result of three factors which reduce our ability to track the halos through the simulation.
Firstly the mass resolution is lower for CAMELS than for TNG100, which means resolution effects are more likely for small halos, and halos which exist at early times in TNG100 might not have enough particles to be identified in CAMELS.
Another factor is that the spacing between snapshots is larger for CAMELS, meaning that it is more difficult to track halos between snapshots as they may have moved further, and gained/lost more mass.
Finally, while we are able to match a high fraction of \textsc{Rockstar} halos with their \textsc{SubFind} counterparts, some halos are matched incorrectly or no match can be found.

Unlike the previous feature importance plots, in Figure \ref{fig:camels_components} each of the components must sum to zero by definition.
This means it is possible to have negative importance values.
This can be seen in the plot of component 1, where dark matter mass always has a negative value.
Consider two simulations, A and B, where simulation A has a larger component 1 value than simulation B, and the coefficients for all the other components are identical.
In this case the relative importance of gas mass to dark matter mass would be greater for simulation A than for simulation B.
This means that increasing component 1 corresponds to a decreasing importance of dark matter mass, which gives the gravitational potential of the halo.
When we increase the speed of supernova winds, they are more likely to blow gas out of the halo.
This is why the dark matter mass is more important for simulations with a large $A_{SN2}$ value, and explains the bifurcation shown in Figure \ref{fig:camels_pca}.

When plotting component 2, we see that there are negative values of gas and dark matter mass at early times, and positive values at late times.
Therefore increasing the component 2 value means that the peak in gas and dark matter mass will occur at a later point in time, corresponding to galaxies which form later.
Increasing $A_{SN2}$ causes galaxies to form later as the gas is ejected further from the galaxy, and so takes longer to cool and return before it can form stars.

For component 3 the difference in gas and dark matter mass is always negative, but is relatively flat.
Thus component 3 corresponds to an increasing importance of the black hole.

\subsection{Comparing parameters}
We now consider the effect of the varying the other simulation parameters.
We have created an online application
\footnote{Available at \href{https://camels-feature-importance.streamlitapp.com/}{https://camels-feature-importance.streamlitapp.com/}}
which estimates the feature importance for a given set of parameters.
The plots shown by the app validate the results found by PCA in the previous section.
For example, if we are predicting stellar mass and we increase the value of the supernova 2 parameter, we see that the dark matter feature importance increases.
To generate the plots for this app we train a Random Forest model that takes simulation parameters as input and predicts feature importance. 

From Figure \ref{fig:camels_tng_coefficients} we can see the effect of modifying the simulation parameters on the feature importance of the CAMELS TNG simulations.
The vertical axis is set to the same scale for all plots to allow for comparison.

The top left panel shows the effect of changing $\Omega_m$.
For low $\Omega_m$ values galaxies form earlier.
It is interesting to compare this with the feature importance plots from the different density environments as shown in Figure \ref{fig:fi_subsamples}.
In that case we found that galaxies in low density regions tended to form later.
Low density regions can be regarded as a separate universe with a low $\Omega_m$ value, so this agrees with our findings from the PCA components.
However, for the different density regions there was no difference in the gas and dark matter mass relative importance, but in CAMELS we do find a correlation with $\Omega_m$.

Changing the value of $\sigma_8$ also effects the time at which galaxies form. Since for low $\sigma_8$ values the density peaks are smaller, it takes longer for the halos to collapse and allow galaxies to form.

Varying $A_{SN1}$, which increases the energy per unit star formation, causes similar but less pronounced effects to $A_{SN2}$. However, it has the opposite effect on the black hole mass relation.

Changing the two parameters associated with the AGN feedback strength has no effect on the mean PCA coefficients.
The reason for this could be because the number of galaxies without supermassive black holes is significantly larger than the number of those with one. 
Ideally we would consider only galaxies above a certain mass cut to exclude those without any black hole activity, but the  $25 (Mpc/h)^3$ box size of the CAMELS simulation is not large enough for us to do this.
However, it also suggests that the black holes in the most massive galaxies are not having any significant effect on the evolution of neighbouring galaxies.

\subsection{Comparing IllustrisTNG and Simba}
We now apply the same analysis to the Simba simulations from the CAMELS suite.
We decompose the feature importance vectors from Simba using the PCA components we obtained from the TNG galaxies.
This allows for a more direct comparison between the two codes.
We remind the reader that due to the different subgrid model implementations the subgrid model parameters (e.g. $A_{SN1}$) have different meaning for TNG and Simba.
See Table \ref{table:camels_params} for a description of each parameter.

In Figure \ref{fig:camels_simba_coefficients} we show the results of this analysis.
In general we find the same trends as in the TNG simulations, but there is often a difference with component 3, which is mainly linked to the black hole importance. 
This is the case for $\Omega_m$, $\sigma_8$, and $A_{SN2}$.

When increasing the value $A_{SN1}$ for Simba we find that galaxies form later, which also occurs for the TNG, but that the halo potential importance decreases, opposite to the TNG.
Tuning the $A_{SN2}$ parameter in the Simba runs produces a larger impact on how late galaxies form than it does for the IllustrisTNG runs.

Unlike for the TNG, changing the black hole feedback parameters does have an impact on the feature importance.
For the bottom right panel we see that decreasing $A_{AGN2}$ means that galaxies form later, due to the fact that their star formation is not being shut down.
However, $A_{AGN1}$ still shows no clear trend with any of the PCA components.

\section{Discussions \& Conclusions}
\label{sec:conclusions}
In this section we add context to our results by discussing how they relate to the existing literature. However, it must be stressed that it is difficult to find direct comparisons with our work. The majority of work comparing different simulations, or examining how well simulations agree with observations, only considers galaxy properties at a single point in time \citep[e.g.][]{2022arXiv221107659A, 2022ApJ...941..205M, 2022MNRAS.516.4084Y}. The method presented in this work is novel in that it makes it possible to determine differences in how properties build up in different simulations, both in terms of the relative importance of physical processes, and the time at which they occur.

When comparing the results of sections \ref{sec:subsamples} and \ref{sec:camels_results} we can sometimes see similar changes to the feature importance, e.g. decreasing the resolution of IllustrisTNG results in a decreased halo mass importance, which is also the result of increasing the supernova strength in CAMELS.
This highlights the number of degeneracies in simulation outputs that can occur from the modelling choices made for a simulation (cosmology, hydrodynamics solver, subgrid models, resolution).
It emphasizes the importance of developing methods that can break these degeneracies. The method shown in this paper can identify what observations can be used to distinguish simulations. For example, Figure \ref{fig:camels_simba_coefficients} shows that varying galactic winds speeds has a larger effect on galaxy formation times than varying the total energy per unit star formation, and so comparisons with observations of galaxy SFHs could be used to calibrate supernova feedback subgrid models.

The standard model of cosmology has proved incredibly effective at providing a good description of a wide range of astrophysical and cosmological data, but there remain observational tensions in the values of cosmological parameters \citep[e.g.][]{2021CQGra..38o3001D, 2022JHEAp..34...49A}. Simulations must take these uncertainties into account. However, it is not clear how much effect the $\Lambda$CDM parameters will have on the evolution of galaxies within a simulation. 
The results shown in this work (which had not before been possible without large data set sizes and data analysis methods) by training a multi-epoch model on the CAMELS simulations show that varying the cosmological parameters has a large effect on the feature importance and therefore on the build up of galaxy properties.
We find that tuning the cosmological parameters has a similar effect size to modifying the subgrid model parameters.
As the majority of recent simulations are not run for a range of cosmological parameters, this needs to be considered before comparing them to observations.
A large number of high-$z$ simulations (e.g. FLARES \citep{2021MNRAS.500.2127L},
Forever22 \citep{2022arXiv221112970Y}, 
SERRA \citep{2022MNRAS.513.5621P}, 
THESAN \citep{2022MNRAS.511.4005K}) have been introduced recently in order to
compare with results from JWST.
Our results are especially relevant in this area since we find a significant effect on how early galaxies form.

Despite being a well researched topic, the impact of resolution on the formation of galaxies is an especially pertinent question currently for two reasons. One is that simulations which explore a large parameter space, such as the CAMELS simulations, cannot be run at high resolution. Another is that upcoming surveys such as Euclid \citep{euclid} and LSST \citep{lsst} are going to cover $\sim$Gp$\text{c}^3$ volumes. Large volume simulations such as the MilleniumTNG \citep{2022arXiv221010060P} and FLAMINGO \citep{flamingo} simulations have been run to compare with these observed volumes, but due to the box size their resolution is low. The results of this work, as shown in Figure \ref{fig:fi_subsamples}, suggest that decreasing resolution by 2 dex has only a minor effect on the formation history of stellar mass. The major times at which star formation occurs is also unchanged. However, this analysis would need to be repeated for a range of resolutions in other simulations than IllustrisTNG.

Differences in hydrodynamical solvers introduce further uncertainty in the galaxy formation.
New solvers continue to be introduced \citep[e.g.][]{2023MNRAS.519..300A, 2023MNRAS.518.4401M}, and tools have been developed to make it easier to compare different methods \citep[e.g.][]{swift}.
Recent work in \cite{2023MNRAS.523.1280B} has shown that different hydrodynamics solvers do not agree even for standard test cases, and thus it is important to consider how they will effect the galaxies produced in cosmological simulations.
Papers \citep[e.g.][]{2015MNRAS.454.2277S, 2019MNRAS.484.2021H} examining the impact of galaxy properties at a single point in time find most are not significantly affected by the details of the hydrodynamics solver.
Both \cite{2014MNRAS.442.1992H} and \cite{2018MNRAS.480..800H} considered the effect of hydrodynamics on the evolution of galaxies, but they ran simulations of isolated objects rather than comparing full cosmological volumes.
Our results consider for the first time the build up of properties in different cosmological simulations, as shown in Figure \ref{fig:fi_different_sims}. 
We have demonstrated that there are more differences in physical drivers of galaxy evolution between IllustrisTNG and Illustris than between IllustrisTNG and EAGLE, showing that the impact of hydrodynamics method is considerably less important than the choice of subgrid models.
Similarly the comparison between TNG300 and TNG100 highlights the ability of subgrid models to address the limitations of resolution.
These results emphasizes the importance of continuing to develop and tune subgrid prescriptions, including well established models such as those used for supernova feedback.

Our conclusions can be summarized as follows:
\begin{itemize}
  \item We have introduced a novel method for extracting information about galaxy formation from simulations by extending the technique from \cite{multi_epoch_1}. A summary of our method is shown in Figure \ref{fig:method_summary}. By considering the feature importance of baryonic properties we can gain insights into the relative importance of different processes and the time at which they occur. We provide a guide for interpreting the resulting plots in Section \ref{subsec:reading_fi_plots}.
  \item In the top row of Figure \ref{fig:fi_subsamples} we examine the impact of resolution. Decreasing the resolution has a clear effect on the feature importance, showing this novel method can be applied as a check for simulation convergence.
  \item From the central row of Figure \ref{fig:fi_subsamples} we can see that galaxies in higher density environments in IllustrisTNG produce stars at earlier times than in low density regions, but the impact of black holes is decreased. We also show that the properties of void galaxies in IllustrisTNG are in agreement with observations.
  \item By directly analysing cosmological simulations we show for the first time that differences due to subgrid models are considerably more significant than those introduced by modelling the gas using \textsc{Arepo} compared with the \textsc{Anarchy} SPH scheme. This can be seen by comparing the feature importance for EAGLE, Illustris, and IllustrisTNG in Figure \ref{fig:fi_different_sims}.
  \item We use PCA to determine the effects of varying the subgrid model parameters within the CAMELS simulations. In Figure \ref{fig:camels_components} we show the feature importance values corresponding to each of the principal components. We find the first component corresponds to the importance of the halo gravitational potential, and the second component relates to the time when galaxy formation takes place.
  \item In Figures \ref{fig:camels_tng_coefficients} and \ref{fig:camels_simba_coefficients} we see that the Simba black hole feedback model has a larger effect on galaxy formation than the IllustrisTNG model, but stellar feedback remains the main driver in both.
  \item Through our analysis of the CAMELS simulations, we discover a substantial dependence between $\sigma_8$ and the time of galaxy formation. Given the current observational tensions in cosmological parameters, it is crucial for high-redshift simulations to consider this aspect when comparing their results with JWST.
\end{itemize}

\section*{Acknowledgements}
The authors are grateful to the anonymous referee for their useful suggestions and questions that helped improved this work.
RM acknowledges support from STFC grant: ST/T506060/1

\section*{Data Availability}

The IllustrisTNG and Illustris data used in this work is available from the IllustrisTNG website (\href{https://tng-project.org}{tng-project.org}).
The CAMELS data is available at the CAMELS website (\href{https://camels.readthedocs.io/en/latest/}{camels.readthedocs.io})
The EAGLE data is available from the their website (\href{https://icc.dur.ac.uk/Eagle}{icc.dur.ac.uk/Eagle}).

The code used to produce this paper is available at \href{https://github.com/robjmcgibbon/multi\_epoch\_2/}{github.com/robjmcgibbon/multi\_epoch\_2}

The analysis in this paper depended on the following packages of the python programming language: NumPy \citep{numpy}, Matplotlib \citep{matplotlib}, Scikit-learn \citep{sklearn}, and ytree \citep{ytree}. We are thankful to the developers of these tools. This research has made intensive use of NASA’s Astrophysics Data System (\href{https://ui.adsabs.harvard.edu}{ui.adsabs.harvard.edu}) and the arXiv eprint service (\href{https://arxiv.org}{arxiv.org}). 

For the purpose of open access, the authors have applied a Creative Commons Attribution (CC BY) licence to any Author Accepted Manuscript version arising from this submission.

\bibliographystyle{mnras}
\bibliography{main}

\appendix

\section{Matching \textsc{Rockstar} and \textsc{SubFind} catalogs}
\label{sec:camels_matching}
We take the positions of subhalos from both the \textsc{Rockstar} and \textsc{SubFind} catalogues. For each of the \textsc{SubFind} halos we locate any \textsc{Rockstar} halos that satisfy the following criteria:
\begin{itemize}
  \item Position within 3x the half mass radius of the \textsc{SubFind} subhalo.
  \item Mass is within a factor 3 of the \textsc{SubFind} subhalo.
\end{itemize}
If there are multiple \textsc{Rockstar} halos which fulfil these criteria we pick the closest one. We repeat this process for every snapshot in the simulation. 

\begin{figure}
	\includegraphics[width=\columnwidth]{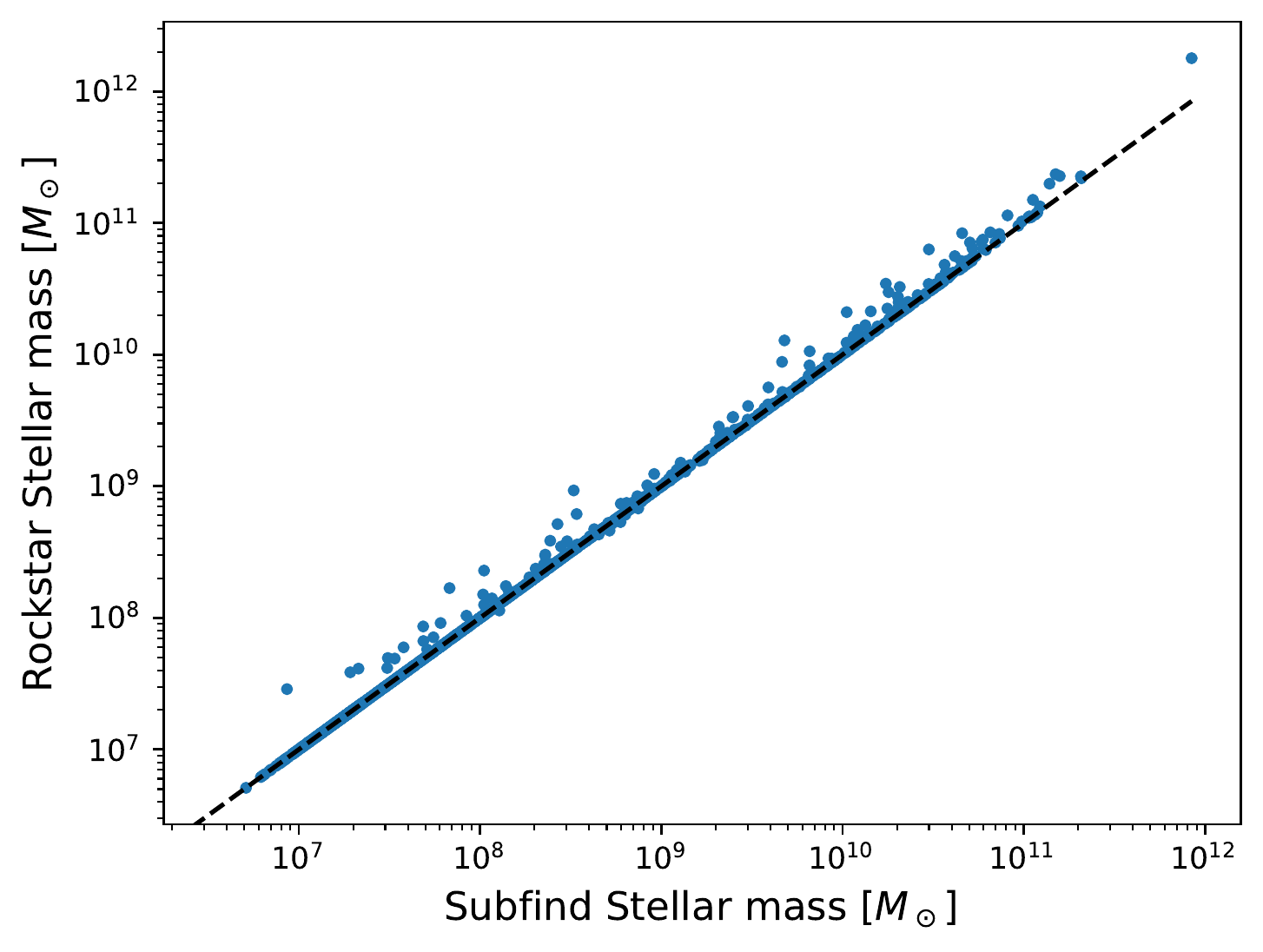}
    \caption{Stellar mass from \textsc{Rockstar} and \textsc{SubFind} catalogues of all the $z=0$ matched subhalos from the IllustrisTNG LH0 simulation. }
    \label{fig:matching_stellar_mass}
\end{figure}

In Figure \ref{fig:matching_stellar_mass} we show the \textsc{Rockstar} and \textsc{SubFind} stellar mass of matched halos.
We have used a larger minimum matching distance than \cite{matching_halos}.
This allows us to gains a larger sample size, while the scatter in the stellar mass as shown in the Figure does not increase significantly.

\section{Model performance}
\label{sec:model_performance}

\begin{figure}
	\includegraphics[width=\columnwidth]{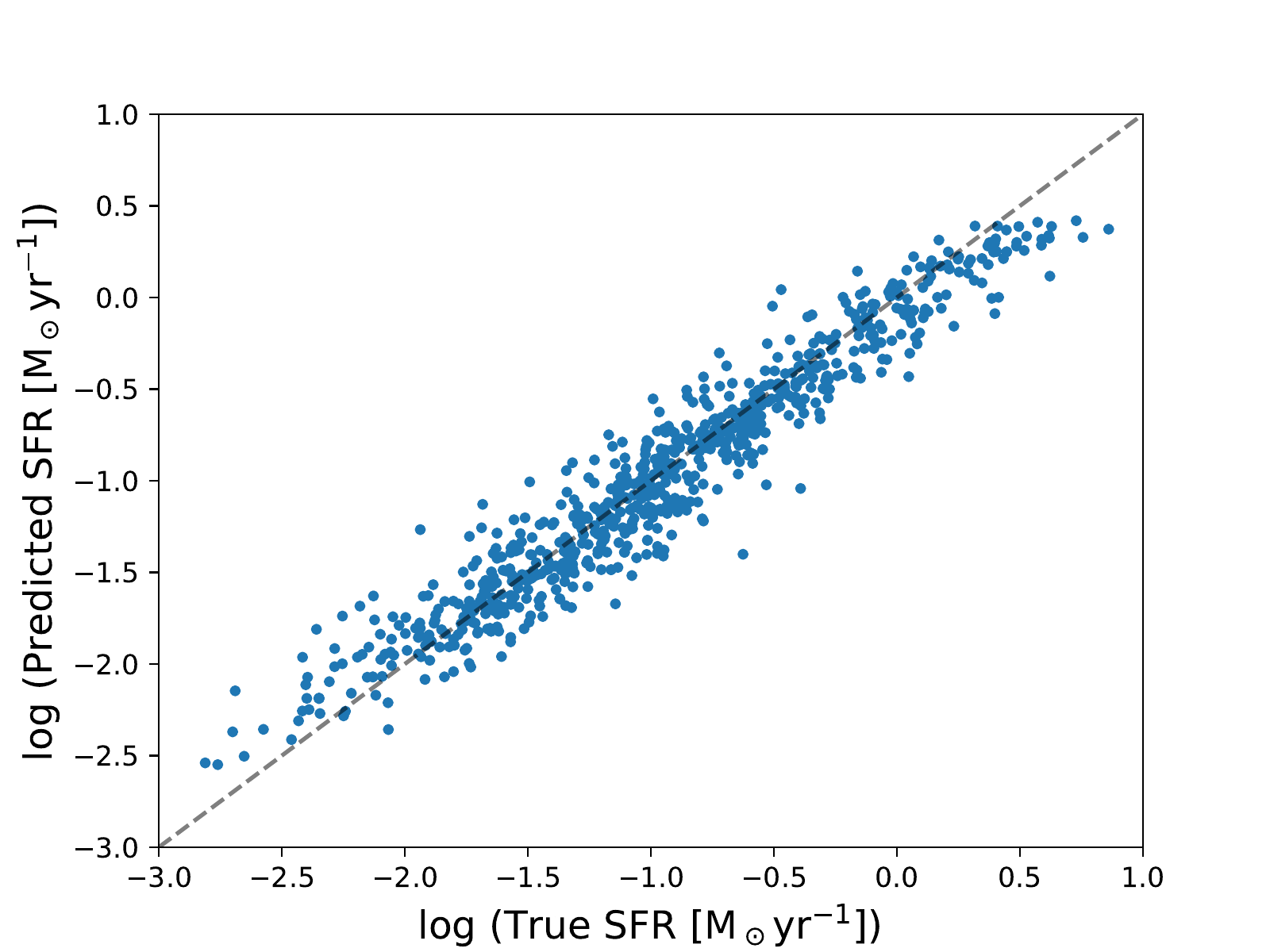}
    \caption{Model performance on 1000 randomly sampled galaxies from the TNG100 test set. The x axis shows the SFR from the simulation, and the y axis shows the value predicted by the machine learning model.}
    \label{fig:sfr_performance}
\end{figure}

The MSE scores from each model are shown in Table \ref{table:model_performance}.
As in \cite{multi_epoch_1} we scale the output values to the range $[0, 1]$ when calculating the MSE. This allows MSE scores from models trained on different data sets to be easily compared. 
As the MSE score for the SFR prediction is significantly worse than for any other model, we show the true vs predicted SFR in Figure \ref{fig:sfr_performance}.

\begin{table}
    \centering
    \caption{The MSE, quantifying the performance of different models at predicting baryonic properties of subhalos. All scores are for predictions on the test set. Values were calculated from averaging 10 train/test splits.}
    \begin{tabular}{|l|l|l|}
        Prediction                          & Figure                            & MSE ($\times10^{-3}$)
        \\ \hline \hline
        TNG100                              & Fig. \ref{fig:fi_tng}             & 0.82
        \\ \hline
        TNG50                               & Fig. \ref{fig:fi_subsamples}      & 0.78
        \\ \hline
        TNG100: Sublink merger trees        & Fig. \ref{fig:fi_subsamples}      & 0.75
        \\ \hline
        TNG300                              & Fig. \ref{fig:fi_subsamples}      & 1.13
        \\ \hline
        TNG100: Low density environment     & Fig. \ref{fig:fi_subsamples}      & 0.90
        \\ \hline
        TNG100: Medium density environment  & Fig. \ref{fig:fi_subsamples}      & 1.08
        \\ \hline
        TNG100: High density environment    & Fig. \ref{fig:fi_subsamples}      & 1.25
        \\ \hline
        TNG100: Stellar metallicity         & Fig. \ref{fig:fi_subsamples}      & 1.16
        \\ \hline
        TNG100: SFR                         & Fig. \ref{fig:fi_subsamples}      & 3.47
        \\ \hline
        TNG100: Gas mass                    & Fig. \ref{fig:fi_subsamples}      & 1.37
        \\ \hline
        TNG100: $z=2$                       & Fig. \ref{fig:fi_different_z}     & 0.51
        \\ \hline 
        TNG100: $z=1$                       & Fig. \ref{fig:fi_different_z}     & 0.82
        \\ \hline
        Illustris                           & Fig. \ref{fig:fi_different_sims}  & 0.64
        \\ \hline
        EAGLE                               & Fig. \ref{fig:fi_different_sims}  & 0.76
        \\ \hline
        EAGLE: FBconst variation            & Fig. \ref{fig:fi_different_sims}  & 0.77
        \\ \hline
        EAGLE: FBZ variation                & Fig. \ref{fig:fi_different_sims}  & 1.01
        \\ \hline
        EAGLE: NoAGN variation              & Fig. \ref{fig:fi_different_sims}  & 0.05
        \\ \hline
    \end{tabular}
    
    \label{table:model_performance}
\end{table}

\bsp	
\label{lastpage}
\end{document}